\documentclass{book}
\usepackage{makeidx,epsfig}
\usepackage{amsthm,amsmath,amssymb}

\usepackage{setspace,graphicx,graphics,color}
\usepackage{Generic}
\usepackage[normalem]{ulem}

\newcommand{\bm}{\boldmath}
\newcommand{\um}{\unboldmath}

\newcommand{\ru}{{\mbox{Uniform}}}
\newcommand{\RR}{{\bf R}}
\newcommand{\cas}{\mbox{${\cal{S}}$}}

\newcommand{\caz}{\mbox{${\cal{Z}}$}}
\newcommand{\beq}{\begin{equation}}
\newcommand{\eeq}{\end{equation}}
\newcommand{\beqn}{\begin{eqnarray}}
\newcommand{\eeqn}{\end{eqnarray}}
\newcommand{\one}{{\mathbf 1}}

\newcommand{\E}{\mbox{E}}

\newcommand{\bW}{\mathbf{W}}

\newcommand{\bo}[1]{\mbox{\bm$#1$\um}}

\numberwithin{equation}{section}
\theoremstyle{plain}

\begin{document}

%

%


%
%
%
%
%

\chapter[Perfect Sampling: A Review]{\centering Perfecting MCMC Sampling: Recipes and Reservations\footnote{This paper is intended for the  Handbook of Markov chain Monte Carlo's $2^{nd}$ edition. The authors will be grateful for any suggestions that could perfect it. }}


\begin{center}
\begin{large}{ Radu V. Craiu \hspace{1em} Xiao-Li Meng}\\
{ radu.craiu@utoronto.ca \hspace{1em} meng@stat.harvard.edu}
\end{large}
\end{center}


\baselineskip=20pt

\section{The Intended Readership}

Research on MCMC methodology  has continued  strong after the publication of this handbook's first edition. We have recorded multiple innovations in making MCMC algorithms more efficient and widely applicable in the context of bigger data and increasingly more complex models. For instance, when likelihoods are expensive to compute and approximations are inserted in the transition kernels, evaluation of incurred errors is done via a class of methods that have come to be known as noisy MCMC (see also chapter ??). One must acknowledge that such developments are in some sense further from perfect sampling than classical MCMC. However, we also recognize a movement in  a direction more aligned  with perfect sampling. Particularly, the emergence of unbiased MCMC as an ingenious use of coupling techniques to eliminate the MCMC bias and to bound the total variation distance to the target (see also chapter ??) promises to change significantly the MCMC landscape in the coming years.   
Since coupling is also the main ingredient in perfect sampling, we introduce the ideas contained in Propp and Wilson's 1996 seminal paper \cite{propp-wilson:exact-sampling} that propagated the general scheme of {\it coupling from the past} (CFTP). Spurred by that idea, there has been intense search and research for perfect sampling or exact sampling algorithms, so named because such algorithms use Markov chains and yet obtain genuine i.i.d.\ draws---hence perfect or exact---from their limiting distributions within a finite number of iterations. 

There is of course no free lunch. Whereas this is a class of very powerful algorithms, their construction and implementation tend to require a good deal of labor and great care. Indeed, even the most basic general themes are not entirely trivial to understand, and subtleties and traps can be overwhelming for  novices. Our central goal for this chapter is therefore to provide an \emph{intuitive} overview of some of the most basic sampling schemes developed since \cite{propp-wilson:exact-sampling}. We do not strike for completeness, nor for mathematical generality or rigorousness. Rather, we focus on a few basic schemes and try to explain them as intuitively as we can, via figures and simple examples. The chapter  is therefore not intended for the residents but rather the visitors of the MCMC kingdom who want  to tour the magic land of perfect sampling. {There are of course a number of other tour packages, including Mark Huber's book on perfect simulation \cite{huber2016perfect}.}
But we hope ours is one of the least expensive ones in terms of readers' mental investment, though by no means are we offering a free ride.

\section{Coupling From the Past (CFTP)}
\label{Vanilla}
\subsection{Moving from Time-Forward to Time-Backward}
The CFTP algorithm is based on an idea that is both simple and revolutionary. Suppose we are interested in sampling from a distribution with probability law $\Pi(\cdot)$ with state space $\cas \subset \RR^d$. We define a Markov chain with stationary law $\Pi$ using a transition kernel $K(x,\cdot)$ whose transitions can be written in a {\it stochastic recursive sequence} (SRS) form
\beq
\label{srs}
X_{t+1}=\phi(X_t,\xi_{t+1}), \; \; t=0, 1, 2, \ldots,
\eeq
where $\phi$ is a deterministic map and $\xi_{t+1}$ is a random variable with state space $\Lambda \subset \RR^r$. (Sometimes  it is automatically assumed that $\Lambda=(0,1)^r$ but that is not  necessary here.) More precisely, the distribution of $\xi$ is such that $P(X_{t+1} \in A)=\Pi(A)=\int K(x_t,A) \Pi(d x_t)$, that is, it guarantees that the output $X_{t+1}$ has the same (marginal) distribution as the input $X_t$ if $X_t\sim \Pi$.

To explain the key idea of CFTP, let us first review the usual implementation of MCMC.
When the chain can be written as in (\ref{srs}), we can simply compute it iteratively starting from an \emph{arbitrary} starting point $X_0\in\cas$, by generating a sequence of $\xi_1, \xi_2, \ldots, \xi_t$, if we decide to run for $t$ iterations.  If the Markov chain formed by (\ref{srs}) is positive recurrent and aperiodic (see Chapter XX), then we know that as $t\rightarrow \infty$,  the probability law of $X_t$, $P_t$, will approach $\Pi$, regardless of the distribution of $X_0$. Of course, how large $t$ needs to be before the difference between $P_t$ and $\Pi$ becomes too small to have practical consequences is the very thorny issue we try to avoid here.

The CFTP, as its name suggests, resolves this problem using an ingenious idea of running the chain \emph{from the past} instead of \emph{into the future}. To see this clearly, compare the following two sequences based on the same random sequence $\{ \xi_1, \xi_2, \ldots, \xi_t\}$ used above:
\beqn
 \mbox{forward: } X_{0\rightarrow t}^{(x)}&=&\phi(\phi( \ldots \phi(\phi(x,\xi_1),\xi_2),\ldots \xi_{t-1}),\xi_t); \label{eq:forward}\\
 \mbox{backward: } {X}_{t\rightarrow 0}^{(x)}&=&\phi(\phi( \ldots \phi(\phi(x,\xi_t),\xi_{t-1}),\ldots \xi_{2}),\xi_1). \label{eq:backward}
\eeqn
The \emph{time-forward} $X_{0\rightarrow t}^{(x)}$ is obviously identical to the $X_{t}$ computed previously with $X_0=x$. The \emph{time-backward} ${X}_{t\rightarrow 0}^{(x)}$  is evidently not the same as $X_{0\rightarrow t}^{(x)}$ but clearly they have the \emph{identical distribution} whenever $\{ \xi_1, \xi_2, \ldots, \xi_t\}$ are exchangeable, which certainly is the case when $\{\xi_t, t=1, 2, \ldots\}$ are i.i.d., as in a typical implementation. (Note a slight abuse of notation, we use $t$ both as the length of the chain and as a generic index.)  Consequently, we see that if we somehow can compute ${X}_{t\rightarrow 0}^{(x)}$ at its limit at $t=\infty$, then it will be a genuine draw from the desired distribution because it has the same distribution as $X_{0\rightarrow t}^{(x)}$ at $t=\infty$.

\subsection{Hitting the Limit}

At  first sight, we seem to have accomplished absolutely nothing by constructing the time-backward sequence because computing ${X}_{t\rightarrow 0}^{(x)}$ at $t=\infty$ surely should be as impossible as computing ${X}_{0\rightarrow t}^{(x)}$ at $t=\infty$! However, a simple example reveals where the magic lies. Consider a special case where $\phi(X_t, \xi_{t+1}) =\xi_{t+1}$, that is, the original $\{X_t, t=1,2, \ldots\}$ already forms an i.i.d.\ sequence, which clearly has the distribution of $\xi_1$ as its stationary distribution (again, we assume $\{\xi_t,  t=1,2, \ldots \}$ are i.i.d.). It is easy to see that in such cases, $X_{0\rightarrow t}^{(x)}=\xi_t$, but $X_{t\rightarrow 0}^{(x)}=\xi_1$, for all $t$. Therefore, with $X_{0\rightarrow \infty}^{(x)}$ we can only say that it has the \emph{same distribution} as $\xi_1$, whereas for $X_{\infty\rightarrow 0}^{(x)}$ we can say \emph{it is} $\xi_1$!

 More generally, under regularity conditions, one can show that there exists a \emph{stopping time} $T$ such that $P(T<\infty)=1$ and that the distribution of $X_{T\rightarrow 0}^{(x)}$ is exactly $\Pi$,
 that is, $X_{T\rightarrow 0}^{(x)}$ ``hits the limit" with probability one. Intuitively, this is possible because unlike $X_t^{(x)}\equiv X_{0\rightarrow t}^{(x)}$, which forms a Markov chain, $\tilde X_t^{(x)}\equiv X_{t\rightarrow 0}^{(x)}$ depends on the entire history of $\{\tilde X_1, \ldots, \tilde X_{t-1}\}$. It is this dependence that restricts the set of  possible paths  $\tilde X_t$ can take and hence makes it possible to ``hit the limit" in a finite number of steps. For a mathematical proof of the existence of such $T$, we refer readers to \cite{propp-wilson:exact-sampling}, \cite{thon-primer} and \cite{tho}.

The CFTP strategy, in a nutshell, is to identify the aforementioned stopping time $T$ via coupling. To see how it works, let us first map $t$ to $-t$ and hence relabel $X_{T\rightarrow 0}^{(x)}$ as $X_{-T\rightarrow 0}^{(x)}$, which makes the meaning \emph{from the past} clearer. That is, there is a chain coming from the infinite past (and hence negative time) whose value at the present time $t=0$ is the draw from the desired stationary distribution. This is because coming from infinite past and reaching the present time is mathematically the same as starting from the present time and reaching the infinite future. However, this equivalence will remain just a mathematical statement if we really have to go into the infinite past in order to determine the current value of the chain. But the fact that the backward sequence can hit the limit in a finite number of steps suggests that for a given infinite sequence $\{\xi_t, t=-1, \ldots \}$, there exists a finite $T$ such that when $t\ge T$, $X_{-t\rightarrow 0}^{(x)}$ will no longer depend on $x$, that is, all paths determined by $\{\xi_t, t=-1, \ldots \}$ will coalesce by time 0, regardless of their origins at the infinite past. It was proved in \cite{foss} that such coupling is possible if and only if the Markov chain $\{X_1, X_2, \ldots \}$ determined by $\phi$ is uniformly ergodic.

\subsection{Challenges for Routine Applications}

Clearly once all paths coalesce, their common value $X_0=X_{-T\rightarrow 0}^{(x)}$ is a genuine draw from the stationary law $\Pi$. Therefore, the CFTP protocol relies on our ability to design the MCMC process given by $\phi$ or more generally by the transition kernel $K$ such that the {\it coalescence of all paths} takes place for moderate values of $T$. This requirement poses immediate challenges in its routine applications, especially for Bayesian computation, where $\cas$ typically contains many states, very often uncountably many. The brute-force way of monitoring each path is infeasible for two reasons. First, it is simply impossible to follow infinitely many paths individually. Second, when the state  space is continuous, even if we manage to reduce the process to just two paths (as with the monotone coupling discussed below), the probability that these will meet is zero if they are left to run independently. Therefore, our first challenge is to design the algorithm so that the number of paths shrinks to a finite one within a few steps. A hidden obstacle in this challenge is being able to figure out exactly {\it which} paths will emerge from this reduction process as they are the ones that need to be monitored until coalescence. The second challenge is to find effective ways to ``force" paths to meet, that is, to couple them in such a way that, at each step, the probability that they take the same value is positive.

The rest of this chapter will illustrate a variety of methods designed to address both challenges and other implementation issues. We do not know any universal method, nor do we believe it exists. But there are methods for certain classes of problems, and some of them are rather ingenious.

\section{Coalescence Assessment}\label{sec:coal}

\subsection{Illustrating Monotone Coupling}

Suppose the space $\cas$ is endowed with a partial order
relationship $\prec$ so that \beq
\label{ord}
x\prec y \Rightarrow \phi(x,\xi)\le \phi(y,\xi)\eeq
 for any $x,y \in \cas$, $\xi \in \Lambda$ and where $\phi$ is an SRS as in (\ref{srs}). If we can find the  minimum and maximum states $X_{\min},X_{\max} \in \cas $ with respect to the order $\prec$, then we can implement this {\it monotone coupler}---as defined by (\ref{ord})---in which it is sufficient to verify the coupling of the paths started at these two extremal points because all other states are ``squeezed" between them. Therefore, the monotone coupler is an efficient way to address the first challenge discussed in Section 1.2.3.  For illustration,  consider the random walk with state space $\cas=\{0.25,0.5,2,4\}$, with probability $p$ moving up or staying if the chain is already at the ceiling state $X_t=4$, and probability $1-p$ moving down or staying if already at the floor state $X_t=0.25$. It is easy to see that this construction forms a monotone chain, expressible as $X_{t}=\phi(X_{t-1}, \xi_t)$, where $\xi_t\sim {\rm Bernoulli}(p)$ and its value determines the direction of the walk, with one going up and zero going down.

\begin{figure}
\begin{center}
\includegraphics[height=4.75in, angle=0]{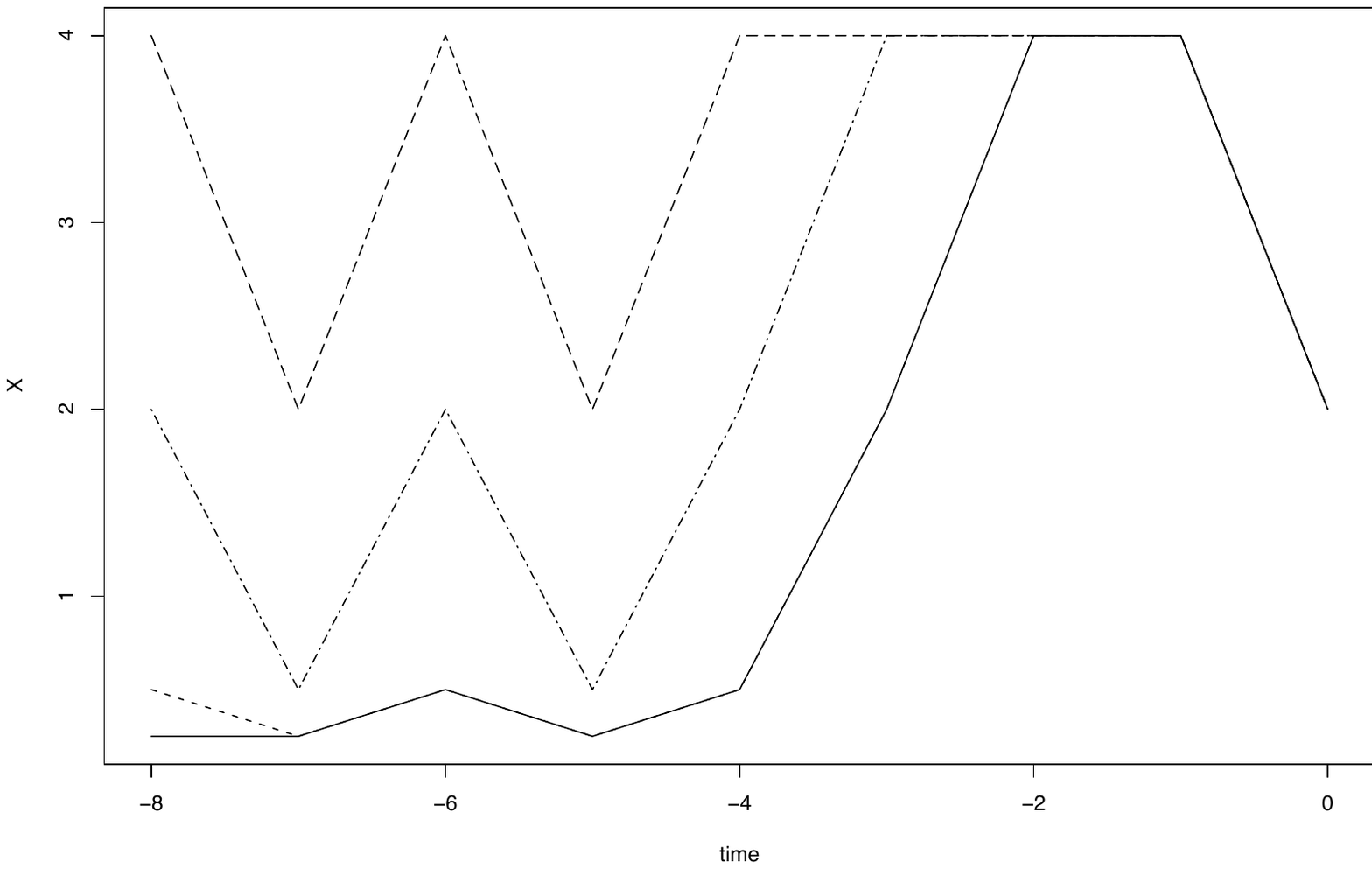}
\caption{\em Illustration of a monotone SRS which preserves the natural order on the real line (i.e., paths can coalesce but never cross each other). Different lines represent sample paths started from different states.}
\label{fig:monot}
\end{center}
\end{figure}

 Figure~\ref{fig:monot} shows a realization of the CFTP process, corresponding to
 \beq\label{eq:seqe}
  \{\xi_{-8}, \xi_{-7}, \ldots, \xi_{-2}, \xi_{-1}\}=\{0, 1, 0, 1, 1, 1, 1, 0\}.
 \eeq
 One can see that the order between paths is preserved by $\phi$. In particular, all the paths are at all times between the paths started at $X_{\min}=0.25$ (solid line) and $X_{\max}=4$ (dashed line), respectively. Therefore, in order to check the coalescence of all four paths, we only need to check if the top chain starting from $X=4$ and the bottom chain starting from $X=0.25$ have coalesced. In this toy example, the saving from checking two instead of all four is obviously insignificant, but one can easily imagine the potentially tremendous  computational savings when there are many states, such as with the Ising model applications in \cite{propp-wilson:exact-sampling}.

\subsection{Illustrating Brute-force Coupling}

This toy example also illustrates well the ``brute-force" implementation of CFTP, that is, checking directly the coalescence of all paths. Figure~\ref{fig:monot} establishes that for any infinite binary sequences $\{\xi_t, t\le -1\}$, as long its last eight values  (i.e., from $t=-8$ to $t=-1$) are the same as that given in (\ref{eq:seqe}), the backward sequence given in (\ref{eq:backward}) will hit the limit $X=2$, that is, the value of the coalesced chain at $t=0$. Pretending that the monotone property was not noticed, we can still check the coalescence step by step for all paths.  Or more efficiently, we can use the ``binary back-off" scheme proposed in \cite{propp-wilson:exact-sampling}, that is, whenever a check fails to detect coalescence, we double the number of ``backward" steps. Specifically, imagine we first made one draw of $\xi$, and it is zero (corresponding to $\xi_{-1}=0$). We compute $X_{-1\rightarrow 0}^{(x)}$ of (\ref{eq:backward}) for all values of $x\in\cas$, which lead to
\beq\label{eq:first}
X_{-1\rightarrow 0}^{(4)}=2, \quad X_{-1\rightarrow 0}^{(2)}=0.5,  \quad X_{-1\rightarrow 0}^{(0.5)}=X_{-1 \rightarrow 0}^{(0.25)}=0.25,
\eeq
indicating that coalescence has not occurred. We therefore double the number of steps going back which  requires only one new draw from $\xi \sim {\rm Bernuolli}(p)$,  as we already have $\xi_{-1}=0$. It is important to emphasize here that we always \emph{reuse} the draws of $\xi_t$'s that we have already made because the point here is to simply check what the coalesced value would be for a \emph{given} infinite sequence of $\{\xi_t, \ t\le -1\}$. The device of making draws starting from $t=-1$ and going backward is the ingenious part of CFTP because it allows us to determine the property of an \emph{infinite} sequence by  revealing and examining only a  \emph{finite} number of its last elements.  That is, since the remaining numbers in  the (infinite) sequence cannot alter the value of the chain at $t=0$, we do not even need to care what they are.

Now this new draw yields $\xi=1$, and hence we have $\{\xi_{-2}, \xi_{-1}\}=\{1, 0\}$, which is then used to compute (\ref{eq:backward}) again but with $T=-2$:
\beq\label{eq:second}
X_{-2\rightarrow 0}^{(4)}=X_{-2\rightarrow 0}^{(2)}=2,  \quad X_{-2\rightarrow 0}^{(0.5)}=0.5, \quad X_{-2\rightarrow 0}^{(0.25)}=0.25,
\eeq
hence, again, no coalescence. Once again we double the steps and go further back to $T=-4$, which means we need two more draws of $\xi$'s, and this time they both are one, yielding $ \{\xi_{-4}, \xi_{-3}, \xi_{-2}, \xi_{-1}\}=\{1, 1, 1, 0\}$. Since we only need at most three consecutive upwards steps to bring any state to the ceiling state $X=4$, the $\{1, 1, 1, 0\}$ sequence immediately implies that
\beq
X_{-4\rightarrow 0}^{(x)}=\phi(4, 0)=2, \quad {\rm for \ all\ } x\in\cas. \nonumber
\eeq
We therefore have detected coalescence after going back to only $T=-4$. This is not in any  contradiction to Figure~\ref{fig:monot}, but points to  an even stronger statement that only the last four elements in the sequence (\ref{eq:seqe}), $ \{\xi_{-4}, \xi_{-3}, \xi_{-2}, \xi_{-1}\}=\{1, 1, 1, 0\}$, not the entire 8 elements, are really relevant.

\subsection{General Classes of Monotone Coupling}
One may wonder when such ordering exists in more general situations and, if so,  what  important classes of distributions can be identified to satisfy (\ref{ord}).
Such questions have been investigated by \cite{haggstrom-nelander:antimonotone}  and \cite{MR2052900} in the case of {\it monotone} (also called {\it attractive})  and {\it anti-monotone} (also called {\it repulsive}) distributions $\Pi$.
Suppose  $\cas=\caz^d$, for some set $\caz \subset \RR$.  We consider the component-wise partial order on
$\cas$ so that $x \prec y$ if and only if $x_i \le y_i$ for all $1\le i \le d$. The probability measure $P$ on $\cas$ is defined to be {\it monotone} if for
each $1\le i\le d$
\beq
\label{eq:mono}
P(X_i \le s|X_{[-i]}=a) \ge P(X_i \le s|X_{[-i]}=b), \ \forall\  s\in\cas
\eeq whenever $a\prec b$ in $\caz^{d-1}$, where
$X_{[-i]}=(X_1,\ldots,X_{i-1},X_{i+1},\ldots,X_d)$.
Similarly, $P$ is called {\it anti-monotone} if \beq
\label{eq:amono}
P(X_i \le s|X_{[-i]}=a) \le P(X_i \le s|X_{[-i]}=b),
\eeq whenever $a\prec b$ in $\caz^{d-1}$.

This definition of monotonicity via all full conditional distributions $P(X_i|X_{[-i]}), i=1,\ldots, d$ was motivated by their use with the Gibbs sampler. In particular, (\ref{ord}) and (\ref{eq:mono}) are easily connected when the sampling from $P(X_i \le s|X_{[-i]}=a)$ is done via the inverse CDF method. Put
$F_i(s|a)=P(X_i \le s|X_{[-i]}=a)$ and assume that the $i$th component is updated using
$\phi_i(x,U)=(x_1,x_2,\ldots,x_{i-1},\inf\{s : \: F_i(s|x_{[-i]})=U\},x_{i+1},\ldots,x_d)$, with $U \sim \ru(0,1)$.  If we assume $x\prec y$, then from (\ref{eq:mono}) we get
\beq\label{eq:monom}
\phi_i(x,U) \prec \phi_i(y,U)
\eeq
because
$\inf\{s: \: F_i(s|x_{[-i]})=U\} \le \inf \{ s : \: F_i(s|y_{[-i]})=U\} $.
Applying (\ref{eq:monom}) in sequential order from $i=1$ to $i=d$, as in a Gibbs-sampler fashion, we can conclude that for $\vec U=\{U_1,\ldots, U_d\}$, the composite map
\beq\label{eq:monof}
\phi(x,\vec U)=\phi_d(\phi_{d-1}(\ldots\phi_2(\phi_1(x, U_1), U_2),\ldots), U_{d-1}), U_d)
\eeq
is monotone in $x$ with respect to the same partial order $\prec$.

In the case of anti-monotone target distributions, it is not hard to see that the $\phi(x, \vec U)$
of (\ref{eq:monof}) is also anti-monotone with respect to $\prec$ if $d$ is odd, but monotone if $d$ is even. Indeed, the ceiling/upper and floor/lower chains switch at each step (indexed by $i=1$ to $i=d$), that is, the ceiling chain becomes the floor chain and vice-versa. This oscillation behavior, however, still permits us to construct \emph{bounding chains} that squeeze in between all the sample paths such that the general coalescence can be detected once the bounding chains have coalesced. See for example \cite{haggstrom-nelander:antimonotone}, which also discusses other examples of monotone target
distributions; also see \cite{crameng2} and \cite{kend}.

\subsection{Bounding Chains}
In a more general setup, \cite{MR2052900} discusses the use of bounding chains without any condition of monotonicity. To better fix ideas, consider the following simple random walk with state space $\cas=\{0.25,0.5,2\}$ and with transition probability matrix (where the $(1,1)$ entry corresponds to the probability that the chain stays at 0.25)
 \[
 A=\left (
 \begin{array}{ccc}
   p &  1-p & 0 \\
  0 & p&  1-p \\
  p & 0 &  1-p\\
  \end{array}
  \right ) .
  \]
Unlike the previous random walk, the recursion defined by the matrix A is neither monotone nor anti-monotone with respect to the natural order on the real line. For example, with $\xi\sim \mbox{Bernoulli}(p)$, and if $\xi=1$, we have $\phi(0.25, \xi)=0.25<\phi(0.5, \xi)=0.5>\phi(2,\xi)=0.25$, where $\phi$ is the chain's SRS. In contrast to the previous random walk, here $\xi=1$ can indicate both moving up or down depending on the starting position, and this is exactly the reason which destroys monotonicity with respect to the same ordering as in the previous random walk example. (This, of course, by no means implies that no (partial) ordering existed under which the SRS is monotone; seeking such an ordering is indeed a common implementation strategy for perfect sampling.)

\begin{figure}
\begin{center}
\includegraphics[height=4.75in, angle=0]{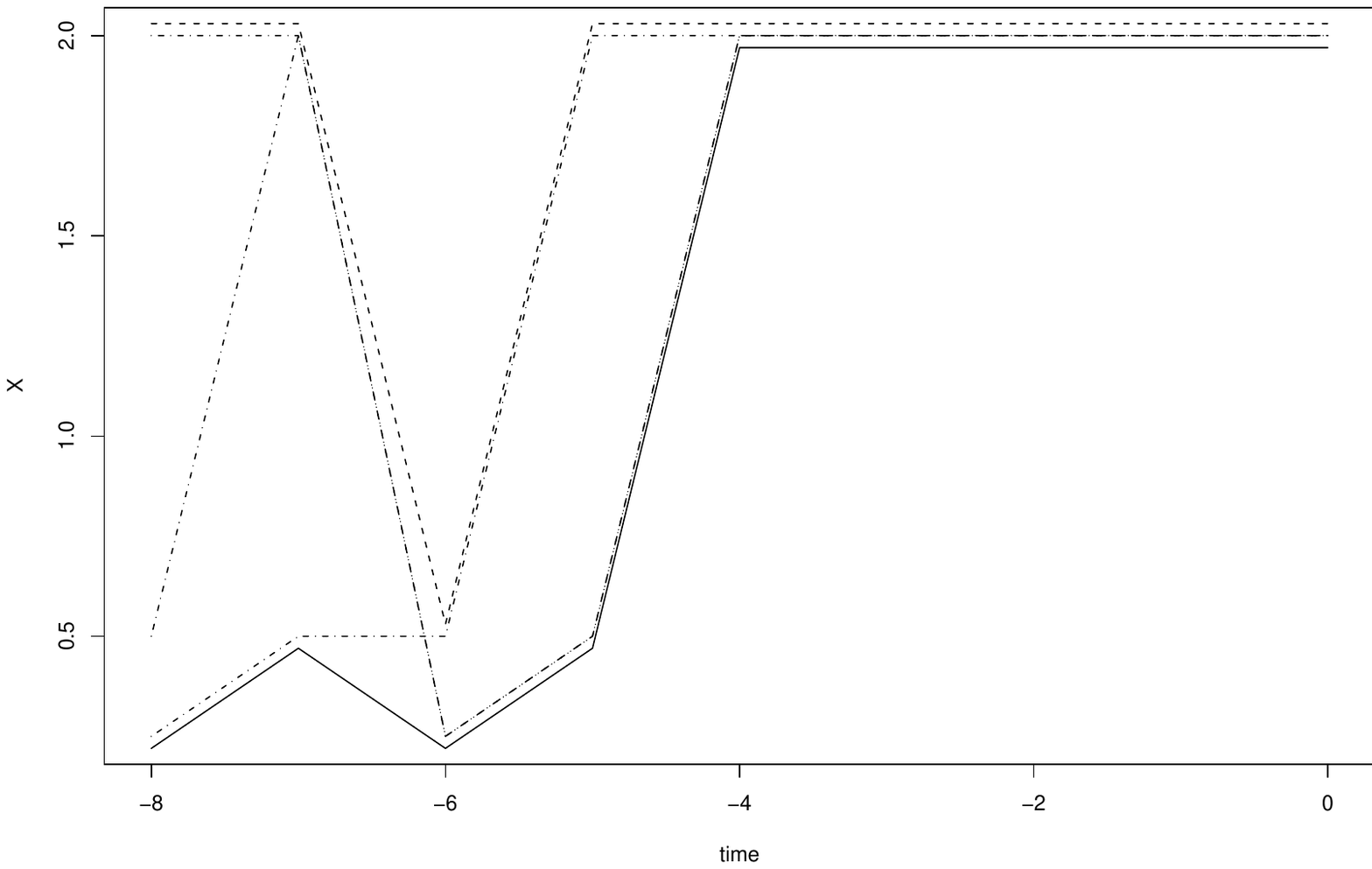}
\caption{\em Non-monotone Markov chain. The dashed and solid lines mark the bounding processes. }
\label{fig:nonmonot}
\end{center}
\end{figure}

 In Figure \ref{fig:nonmonot} we show one run of the CFTP algorithm implemented for this simple example with $p=0.1$, where $\{\xi_{-8}, \ldots, \xi_{-1}\}=\{ 0, 1, 0, 0, 0, 0, 0, -1\}$. One can see that the three paths cross multiple times and no single path remains above or below all the others at all times.  A bounding chain, in the general definition introduced by \cite{MR2052900}, is a chain $\{Y_t   : \; t \ge 0\}$ defined
 on $2^ {\cas} $, the set of all subsets of $\cas$, with the property that if $X_t^{(x)} \in Y_t$ for all $x \in \cas$ then $X_{t+1}^{(x)} \in Y_{t+1}$ for all $x\in \cas $; evidently $Y_0$ needs to contain all values in $S$.  If, at some time $t$, $Y_t$ is a singleton then coalescence has occurred. Clearly, there are many ways to define the chain $Y_t$ but only a few are actually useful in practice and these are obtained, usually, from a  careful study of the original chain $X_t$.

 For instance, in our example we notice that after one iteration $Y_0=\cas$ will become
 either $Y_1=\{0.25, 0.5\}$ or $Y_1=\{0.5, 2\}$, depending on whether $\xi=1$ or $\xi=0$, and therefore $Y_t$ will always be a subset of these two sets (possibly themselves). Therefore, for $t\ge 1$, the updating rule $Y_{t+1} = \Psi(Y_t,\xi)$ can be simplified to
 \beq\label{eq:bond}
 \Psi(Y_t,\xi)= \left \{
 \begin{array}{cc}
Y_t, & \mbox{ if  }\xi =1 \mbox{ and } Y_t = \{0.25,0.5\}; \\
\{0.25,0.5\} &  \mbox{ if  }\xi =1 \mbox{ and } Y_t = \{0.5,2\};\\ 
 \{0.5,2\}, & \mbox{  if  } \xi=0 \mbox{ and } Y_t=\{0.25,0.5\};\\
 \{2\}, & \mbox{ if } \xi=0 \mbox{ and  } Y_t=\{0.5,2\};\\
 \phi(X_t,\xi), & \mbox{ if  } Y_t=\{X_t\}.\\
 \end{array}
\right .
\eeq
One can see then that having the ordered triplet  $\{1,0,0\}$ in the $\xi$-sequence triggers coalescence after which one simply follows the path to time zero.

Two essential requirements for an effective bounding chain are that (I) it can detect coalescence of the original chain and (II) it requires less effort than running all original sample paths. The chain $Y_t\equiv \{ \cas\}$ for all $t$ is a bounding chain and satisfies (II), but clearly it is useless. As an example of bounding chains that do not satisfy (II), consider the dashed path and solid path in Figure \ref{fig:nonmonot}. Here the dashed path is the maximum value attained by all paths at each time $t$, and the solid path is for the minimal value (both have been slightly shifted for better visualization). For each time $t$, the interval between the dashed and the solid paths, denoted by
$\tilde{Y}_t$, clearly forms a bounding chain. But unlike $Y_t$ of (\ref{eq:bond}), the updating function for $\tilde Y_t$ is not easy to define so running $\tilde Y_t$ involves checking the extremes of all the paths for $X_t$ and is, thus, as complicated as running all paths for $X_t$.

As far as general strategies go, \cite{haggstrom-nelander:antimonotone} shows how to construct bounding chains when each component of the random vector $X$ is updated via a Gibbs sampler step, whereas
\cite{MR2052900} presents a general method for constructing bounding chains and applies it to problems from statistical mechanics and graph theory. From the time of this chapter's  initial publication, \cite{stein2013practical} provides a composition strategy of alternating between a monotonic and an anti-monotonic perfect sampling algorithm to significantly improve the computational efficiency.  It is also an example of using discrete data augmentation to achieve effective coalescence, a strategy we shall illustrate in Section~\ref{coup-aug}. 
~\section{Cost-saving Strategies for Implementing Perfect Sampling}

The vanilla CFTP described in Section \ref{Vanilla} suffers from two main drawbacks. First, the implementation ``from the past" requires storing the random seeds used in the backward process  until coupling is observed and a random sample is obtained.  Second, the impatient user cannot abandon runs that are too long without introducing sampling bias, because the coupling time $T$ is correlated with the sample obtained at time zero. In the following two sections we provide intuitive explanations of the read-once CFTP and Fill's interruptible algorithm, designed respectively to address these two drawbacks.

\subsection{Read-once CFTP}\label{sec:readonce}
\begin{figure}
\begin{center}
\includegraphics[width=\textwidth]{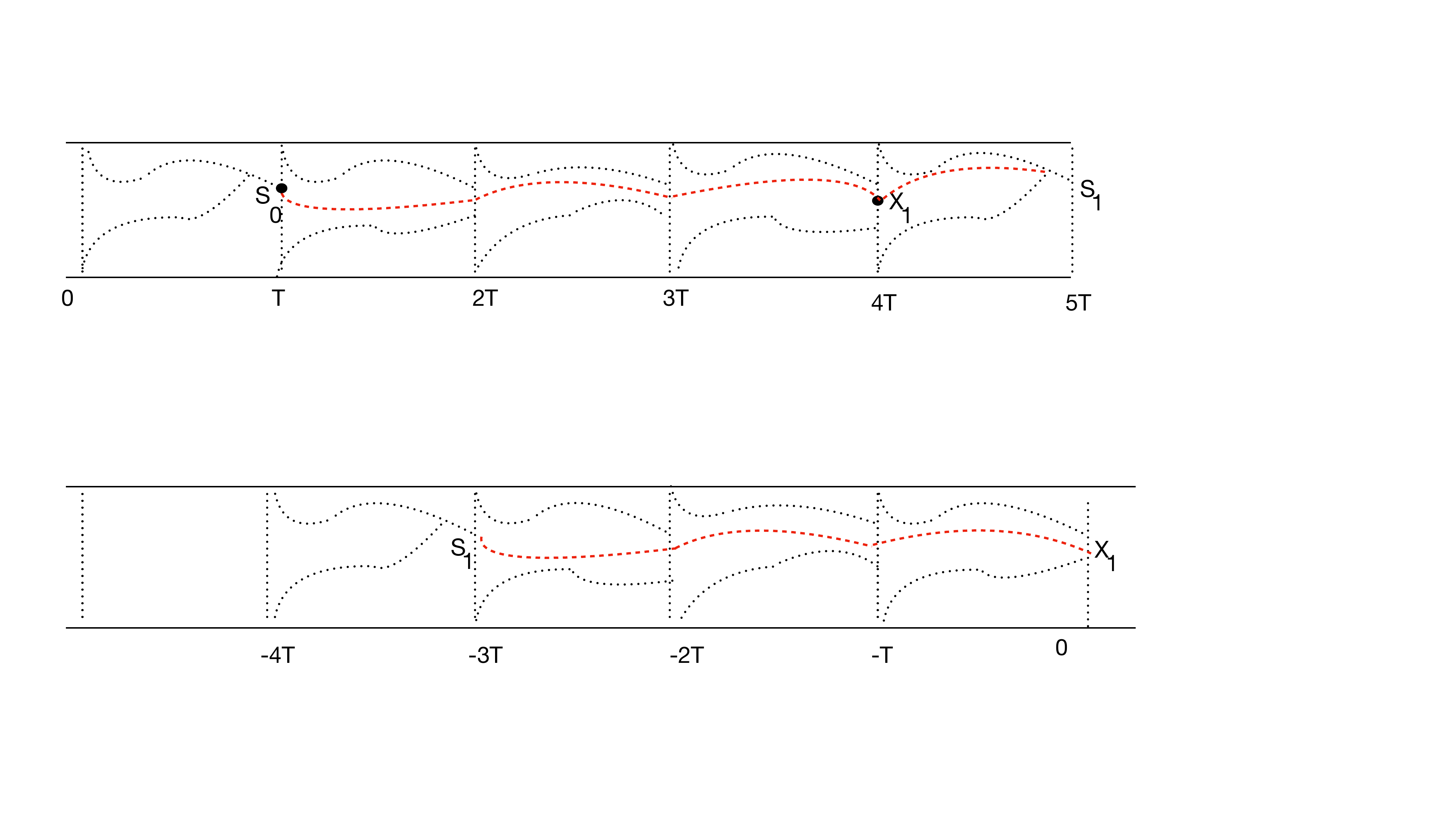}
\caption{\em Top: The read-once CFTP with blocks of fixed length.
Bottom: Comparison with CFTP2.}
\label{fig:fbro}
\end{center}
\end{figure}

Read-once CFTP (Ro-CFTP) as proposed by Wilson (\cite{read-once}) is a clever device that turns CFTP into an equivalent ``forward-moving" implementation. It collects the desired i.i.d.\ draws as the process moves forward and without ever needing to save any of the random numbers previously used. The method starts with a choice of a fixed block size $K$, such that the $K$-composite map
\begin{equation}
\phi_K(x; \vec{\xi})= \phi(\phi(\ldots \phi(\phi(x,\xi_1),\xi_2),\ldots, \xi_{K-1}), \xi_K),
 \end{equation}
 where $\vec{\xi}=\{\xi_1, \ldots, \xi_K\}$, has a high probability to coalesce, that is,
 the value of $\phi_K(x; \vec{\xi})$ will be free of $x$, or equivalently, all paths coalesce \emph{within the block} defined by $\vec{\xi}$. In \cite{read-once}, it is suggested  to select  $K$ such that the probability for $\phi_K$ to coalesce, denoted by $p_K$, is at least 50\%. Given such a $\phi_K$, we first initialize the process by generating i.i.d.\ $\vec{\xi}_j, j=1,2, \ldots$ until we find a $\vec{\xi}_{j_0}$ such that $\phi_K(x; \vec{\xi}_{j_0})$ coalesces.  Without loss of generality, in the top panel of Figure~\ref{fig:fbro}, we assumed that $j_0=1$; and we let $S_0=\phi_K(x; \vec{\xi}_{j_0})$. We then repeat the same process, that is, generating i.i.d.\ $\vec{\xi}_j$'s until $\phi_K(x; \vec{\xi})$ coalesces again.
 In the top panel of Figure~\ref{fig:fbro}, this occurred after four blocks. We denote the coalescent value as $S_1$. During this process, we follow from block to block only the {\it coalescence path} that goes through $S_0$ while all the other paths are reinitialized at the beginning of each block. The location of the coalescence path {\it just before} the beginning of the next coalescent composite map is a sample from the desired $\Pi$.  In Figure~\ref{fig:fbro} this implies that we retain  $X_1$ as a sample. The process then is repeated as we move forward, and this time we follow the path starting from $S_1$ and the next sample $X_2$ (not shown) is the output of this path immediately before the beginning of the next coalescent block. We continue this process to obtain i.i.d.\ draws. 

The connection between the Ro-CFTP and CFTP may not be immediately clear. Indeed, in the vanilla CFTP, the concept of a composite/block map is not emphasized because, although we ``back-off" in blocks, we do not require to have a coalescent composite map of fixed length. For instance, if we set $K=4$, we can see that in Figure \ref{fig:monot} the paths started at $-2K$ coalesce in the interval $(-K,0)$ rather than within the block $(-2K, -K)$.
However, suppose we consider a modified implementation of the vanilla CFTP, call it CFTP2, in which we go back from time zero block by block, each with size $K$, until we find a block that is coalescent, that is, all paths coalesce \emph{within that block}. Clearly, if we trace the path from the coalescent value from that block until it reaches time zero, it will be exactly the same value as found by the original vanilla CFTP because once the coalescence takes place, all paths will stay together forever. {The bottom panel of Figure~\ref{fig:fbro} illustrates CFTP2, where the fourth block (counting backward from $t=0$) is the coalescent block, and $X_1$ is our draw.} 

{Note that the probability of coalescence within any given block is the same, implying that the number of blocks between $S_0$ and $S_1$, say $J$, is a random variable that has a geometric distribution. Figure~\ref{fig:fbro}  shows that Ro-CFTP yields a path in which once $S_0$ is recorded, the sample $X_1$ is collected after $J-1$ blocks (in this case $J=4$). The bottom panel illustrates that the corresponding CFTP2 sample is obtained from a path in which the first coalescent block is the $J$-th one from the origin. We  prove  below that  after a coalescent block, it is most efficient to update the path for $J-1$ blocks before recording a sample from the target.  } 

The resemblance of the bottom panel and the first three blocks in the top panel (counting
forward from time $t=0$) is intended to highlight the equivalence between Ro-CFTP and CFTP2. On its own, CFTP2 is clearly less cost-effective than CFTP because by insisting on having block coalescence, it typically requires going back further in time than does the original CFTP (since block coalescence is a more stringent detecting criterion, as discussed above). However, by giving up a bit on the efficiency of detecting coalescence, we gain the \emph{independence} between the block coalescent value $S_0$ and the entire backward search process for $S_0$ and hence we can reverse the order of the search without affecting the end result.

As this independence is the backbone of the Ro-CFTP, here we show how critically it depends on having \emph{fixed-size} blocks. Intuitively, when the blocks all have the same size, they all have the same probability to be a coalescent block, and the distribution of the coalesced state given a coalescent block is the same regardless which block it is. To confirm this intuition and see how it implies the independence, let us define the block random vector $\vec{\xi}_{-t}=(\xi_{-tK},\xi_{-tK+1}\ldots, \xi_{-tK+K-1})$ and, for a given set $\vec{\xi}_{-t}, t=1, 2, \ldots$, let $T$ be the first $t$ such that $\phi_K(x; \vec{\xi}_{-t})$ coalesces, and let $S_0=\phi_K(x; \vec{\xi}_{-T})$ be the coalescent value. Also let $C_j=\{ \phi_K(x, \vec{\xi}_{-j})\ coalesces\}$, that is, the event that the $j$th block map coalesces. Then $\{T=t\}=(\cap_{j=1}^{t-1}C_j^c) \cap C_t $. For notational simplicity, denote $A_j=\{\phi_K(x, \vec{\xi}_{-j})\in A\}$ and $B_j=\{\Xi_{j}\in B\}$, where
$A$ and $B$ are two arbitrary (measurable) sets on the appropriate probability spaces, and $\Xi_j=\{\vec\xi_{-1},\ldots, \vec\xi_{-j}\}$.
Then for any positive integer $t$,
\begin{eqnarray}\nonumber
& & P(S_0\in A, T=t, \Xi_{T-1}\in B) =P(A_t\cap[\cap_{j=1}^{t-1}C_j^c \cap C_t ]\cap B_{t-1})\\
\label{eq:indep}
&=& P(A_t\cap C_t)P(\cap_{j=1}^{t-1}C_j^c\cap B_{t-1})= P(A_t|C_t)P(C_t)P(\cap_{j=1}^{t-1}C_j^c\cap B_{t-1})\\ \nonumber
&=& P(A_t|C_t)P(C_t\cap_{j=1}^{t-1}C_j^c \cap B_{t-1})= P(A_1|C_1)P(T=t, B_{T-1}).
\end{eqnarray}
In deriving the above equalities, we have repeatedly used the fact that $\{A_t, C_t\}$ are independent of $\{A_{t-1}, B_{t-1}, C_{t-1}\}$ since they are determined respectively by $\vec{\xi}_{-t}$ and $\{\vec{\xi}_j, j=-1, \ldots, -(t-1)\}$. The last switching from
$P(A_t|C_t)$ to $P(A_1|C_1)$ is due to the i.i.d.\ nature of $\{A_t, C_t\}$'s, because all blocks have the same size $K$. This switching is critical in establishing the factorization  in (\ref{eq:indep}), and hence the independence.

Clearly, as depicted in Figure~\ref{fig:fbro}, the output of CFTP2, namely $X_1$,
can be expressed as $M(S_0, T, \Xi_{T-1})$, where $M$ is a \emph{deterministic} map. The aforementioned independence ensures that if we can find $\{\tilde T, \Xi_{\tilde T-1}\}$ such that it has the same distribution as $\{T, \Xi_{T-1}\}$ and is independent of $S_0$, then $\tilde X_1=M(S_0, \tilde T, \Xi_{\tilde T-1})$ will have the same distribution as $X_1=M(S_0, T, \Xi_{T-1})$, and hence it is also an exact draw from the stationary distribution $\Pi$. Because $\{\vec{\xi}_{-1}, \vec{\xi}_{-2}, \ldots, \}$ are i.i.d., obviously the distribution of $\{T, \Xi_{T-1}\}$ is invariant to the order at which we check for the block coalescence. We therefore can reverse the original backward order into a  forward one and start at an arbitrary block which must be independent with $S_0$. This naturally leads to the Ro-CFTP, because we can start with the block immediately after a coalescence has occurred (which serves as $S_0$), since it is independent of
$S_0$. 

Moreover,  the number of blocks  and all the block random numbers (i.e., $\xi$'s) needed \emph{before} we reach the next coalescent block   represents  a sample from the distribution of $\{T, \Xi_{T-1}\}$. (It is worth emphasizing that each coalescent composite map fulfills two roles as it marks the end of a successful run (inclusive) and the beginning of a new run (exclusive).) Alternatively, (\ref{eq:indep}) implies that we can first generate $T$ from a geometric distribution with mean $1/p_K$ (recall $p_K$ is the probability of coalescence within each block), and then generate $T-1$ non-coalescent blocks, via which we then run the chain forward starting from $S_0$. This observation has little  practical impact since $p_K$ is usually unknown, but it is useful for understanding the connection with the splitting chain technique that will be discussed in Section \ref{coupmet}. The forward implementation brought by Ro-CFTP makes it also easier to implement the efficient use of perfect sampling tours proposed by \cite{eff-cftp}, which will be discussed in Section \ref{swind}.

\subsection{Fill's algorithm}\label{sec:fill}

Fill's algorithm \cite{fill} and its extension to general chains \cite{fill:machid} breaks the dependence between the backward time to coalescence and the sample obtained at time zero.  In the  following we use the slightly modified  description from \cite{jeff-dunc}.

The algorithm relies on the time reversal version of the Markov chain designed to sample from $\Pi$.  If the original chain has transition kernel $K(x, \cdot)$, then the time reversal version has kernel $\tilde{K}(z, \cdot)$, such that
\beq
\label{eq:db}
\tilde{k}(x|z)\pi(z)=k(z|x)\pi(x),\quad \forall \ (x, z) \in \cas\times\cas,
\eeq
where for simplicity of presentation, we have assumed the stationary law $\Pi$ has density $\pi$, and $K(x,\cdot)$ and $\tilde{K}(z,\cdot)$ have kernel densities $k( \cdot |x)$ and $\tilde{k}(\cdot |z)$ respectively.
It also requires that given a particular path $X_0 \rightarrow X_1 \rightarrow \ldots \rightarrow X_t$, we can sample, {\it conditional on the observed path}, a sample of the same length from any state in $\cas$.

\begin{figure}
\begin{center}
\includegraphics[height=6in]{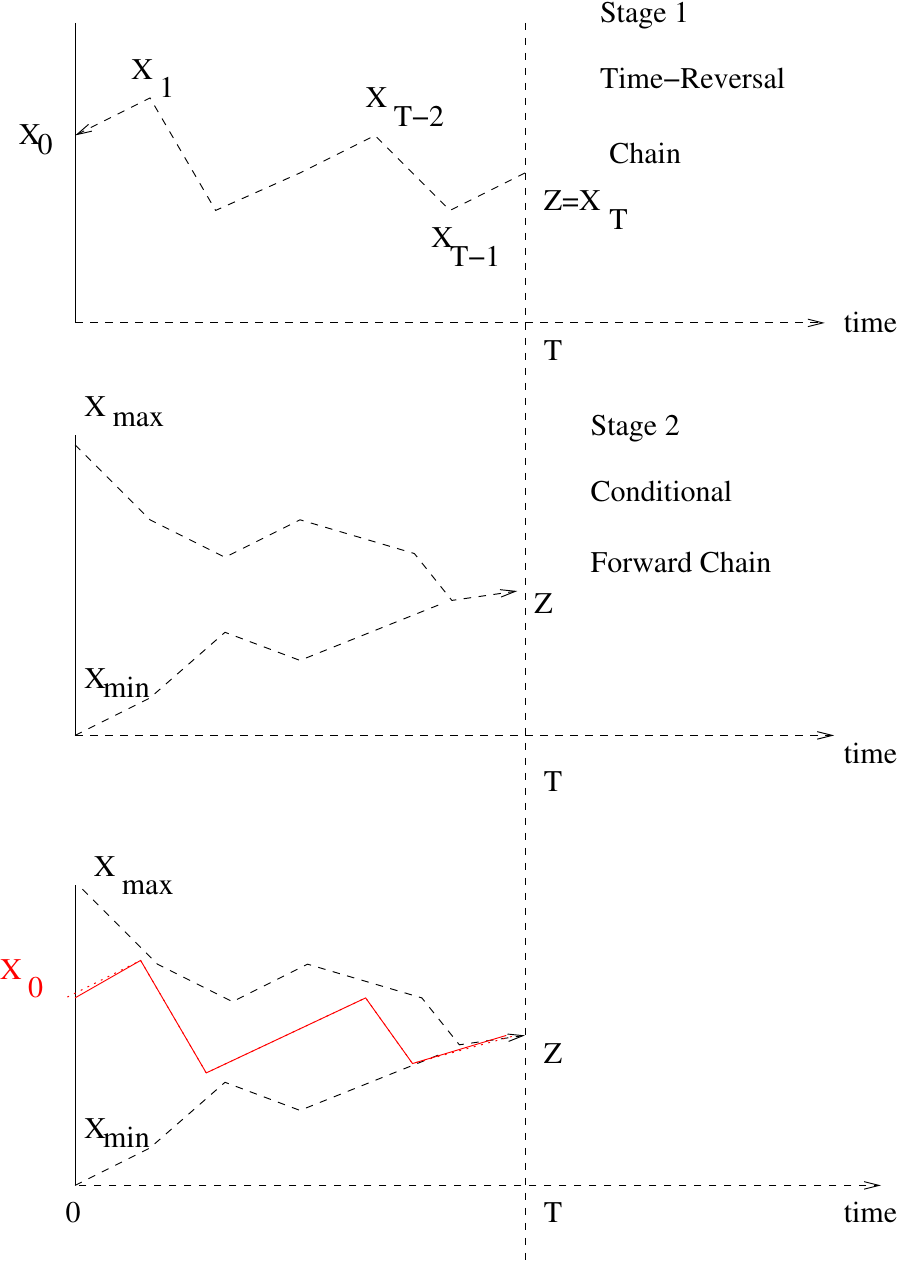}
\caption{{\em Illustration of Fill's algorithm }}
\label{fig:fill}
\end{center}
\end{figure}

The algorithm starts by sampling a random $Z \in \cas$ from an arbitrary distribution $P_0$ (with density $p_0$)
that is absolutely continuous with respect to $\Pi$, and by selecting a positive integer $T$.
 The first stage  is illustrated in the top panel of Figure \ref{fig:fill}:  using the reversal time chain we  simulate  the path
$Z=X_T \rightarrow X_{T-1} \rightarrow \ldots \rightarrow X_1 \rightarrow X_0$ (note that the arrow is pointing against the time's direction). In the second stage, we sample forward from all the states in $\cas$ conditional on
the existing path  $X_0 \rightarrow X_1 \rightarrow \ldots \rightarrow X_T=Z$ (note that this path is considered now in the same direction as time). The conditional sampling is the main difficulty encountered when implementing Fill's algorithm because while the initial run of the chain is done in the time-reversed order (from $X_T=Z$ to $X_0$), the conditioning is done on the path defined by that initial run that is now considered in the chronological order (from $X_0$ to $Z$). For instance, suppose one runs the former using recursion $X_{t-1}=\phi_{\leftarrow}(X_t, V_{t-1})$ and the latter using $X_t=\phi_{\rightarrow}(X_{t-1},U_t)$.  If the chain $\{X_t\}$ is reversible then $\phi_\leftarrow=\phi_\rightarrow$  but one must still determine the  distribution of the sequence 
$\{ U_t: \; 0\le t \le T\}$ conditional on yielding  the path $X_0 \rightarrow X_1 \ldots \rightarrow X_T=Z$. 
 When the chain is non-reversible, one must also determine $\phi_{\rightarrow}$ from $\phi_\leftarrow$.



To illustrate, consider the multimodal target that was introduced in \cite{gelman1991note}:
\[\pi(x,y) \propto \exp \left (-\frac{8x^2y^2+x^2+y^2-4xy-8x-8y}{2} \right).\]
Sampling from $\pi$ can be done via a Gibbs sampler that consists of two steps:  
\begin{eqnarray}
x_{t+1}|y_t &\sim& \mbox{N}\left(\frac{2y_{t}+4}{8y_{t}^2+1}, \frac{1}{8y_{t}^2+1}\right); \label{s1}\\
y_{t+1}|x_{t+1} &\sim& \mbox{N}\left(\frac{2x_{t+1}+4}{8x_{t+1}^2+1}, \frac{1}{8x_{t+1}^2+1}\right).\label{s2}
\end{eqnarray}
For each step of the forward chain one can define a recursion function $\phi_{\rightarrow}((x_t,y_t);u_{t+1},w_{t+1})$. Updating $x$ with \eqref{s1} corresponds to 
\beq
x_{t+1}=\phi_{x, \rightarrow}((x_{t}, y_{t}); {u}_{t+1})= \frac{2y_{t}+4}{8y_{t}^2+1}+ u_{t+1} \left( \frac{1}{8y_{t}^2+1}\right )^{1/2} \label{srs-x}
\eeq
while \eqref{s2} is done via 
\beq
y_{t+1}= \phi_{y, \rightarrow}((x_{t+1}, y_{t}); {w}_{t+1})=\frac{2x_{t+1}+4}{8x_{t+1}^2+1}+w_{t+1}\left( \frac{1}{8x_{t+1}^2+1}\right )^{1/2}, \label{srx-y}
\eeq
 where   ${u}_{t+1}$ and  ${w}_{t+1}$
are iid $\mbox{N}(0,1)$ random variables.
Putting these two steps together, we have 
\beqn
(x_{t+1},y_{t+1})&=&\phi_\rightarrow((x_t,y_t);u_{t+1},w_{t+1}) 
= \phi_{y, \rightarrow}((\phi_{x, \rightarrow}((x_{t}, y_{t}); {u}_{t+1}),y_t);w_{t+1}).
\label{reverso}
\eeqn
 In this case, the chain's transition kernel is not reversible. However,  the reverse chain is run simply by reversing the order of updates so
that 
\beqn
(x_t,y_t) = \phi_{\leftarrow}((x_{t+1},y_{t+1}); {v}_{t}, {z}_{t}) =
\phi_{x,\rightarrow}( (x_{t+1},\phi_{y,\rightarrow}((x_{t+1},y_{t+1});{z}_{t})),{v}_{t}), \label{reverse}
\eeqn
where 
$v_t$ and $z_t$ are iid $\mbox{N}(0,1)$.

To implement Fill's algorithm, we first need run the reverse chain to obtain a path $(x_T,y_T) \rightarrow (x_{T-1},y_{T-1}) \rightarrow\ldots (x_1,y_1)\rightarrow (x_0,y_0).$ 
In the second stage, we need to run the chain in chronological order  conditional on ``hitting"  $(x_1,y_1) \rightarrow \ldots \rightarrow 
 (x_T,y_T)$ with the path that is started in $(x_0,y_0)$. 
 
This implies finding the conditional distribution of $(u_1,w_1),\ldots (u_T,w_T)$, given the forward  trajectory (which is itself imposed by the reverse chain run). In this case, the conditional distribution of each $(u_i, w_i)$ is a Dirac measure with mass in a point characterized by $(x_i,y_i)$ and $(x_{i-1},y_{i-1})$. For example, the random deviates that will transition the forward chain from $(x_0,y_0)$ to $(x_1,y_1)$ must satisfy
\beq
u_1= \left (x_1 - \frac{2y_{0}+4}{8y_{0}^2+1}\right) \left( \frac{1}{8y_{0}^2+1}\right )^{-1/2} \nonumber
\eeq
and
\beq
w_1=\left ( y_1  - \frac{2x_{1}+4}{8x_{1}^2+1} \right) \left( \frac{1}{8x_{1}^2+1}\right )^{-1/2}.\nonumber
\eeq
Note that $u_1,w_1$ are different from $v_0,z_0$ since the latter satisfy a slightly different set of conditions
\beq
v_0=\left ( y_0  - \frac{2x_{1}+4}{8x_{1}^2+1} \right) \left( \frac{1}{8x_{1}^2+1}\right )^{-1/2}
\nonumber
\eeq
and
\beq
z_0=\left ( x_0  - \frac{2y_{0}+4}{8y_{0}^2+1} \right )\left( \frac{1}{8y_{0}^2+1}\right )^{-1/2}. 
\nonumber
\eeq
Unfortunately, when the forward recursion is more complex, identifying the conditional distribution of the random variates $\{(u_t,w_t): \; 1\le t \le T\}$  can be very difficult, thus  hindering the broad application of Fill's algorithm.

If by time $T$ all the paths have coalesced, as depicted in the middle panel of Figure \ref{fig:fill} (where we used monotone
coupling for simplicity of illustration, but the idea is general), we retain $X_0$ as a sample from $\pi$, as shown in the bottom panel of Figure \ref{fig:fill}, and restart with a new pair $(Z,T)$.
Otherwise, we select a new $T$ or we restart with  a new pair $(Z,T)$.

To understand why the algorithm produces i.i.d.\ samples from $\pi$, we first note that (\ref{eq:db}) holds in the more general form
\beq
\label{eq:gdb}
\tilde{k}_t(x|z)\pi(z)=k_t(z|x)\pi(x),\quad \forall\  (x, z)\in \cas\times\cas,
\eeq
where $k_t$ is the kernel density of the $t$-step forward transition kernel $K_t$ and $\tilde{k}_t$ is for the corresponding time reversal one, $\tilde{K}_t$. Fill's algorithm retains only those paths from $Z$ to $X_0$ (obtained via $\tilde K_T$) such that the corresponding $k_T(z|x)$ is free of $x$---and hence it can be expressed as $h_T(z)$---due to coalescence; in this sense Fill's algorithm is a case of rejection sampling. Therefore, using (\ref{eq:gdb}) the probability density for those retained $X_0$'s is
\beq\label{eq:fill}
p(x)=\int\tilde{k}_T(x|z)p_0(z)dz= \int \frac{\pi(x) h_T(z)}{\pi(z)} p_0(z)dz \propto \pi(x),
 \eeq
 and hence the correctness of the sampling algorithm (see also \cite{clr}). Note that here, for simplicity, we have deliberately blurred the distinction between the fixed $t$ in (\ref{eq:gdb}) and potentially random $T$ in (\ref{eq:fill}); in this sense (\ref{eq:fill}) is  a heuristic argument for building intuition rather than a rigorous mathematical proof.   In its general form, Fill's algorithm can search for the coalescence time $T$ just as with CFTP---see \cite{fill:machid} for a detailed treatment of the general form of Fill's algorithm, including a rigorous proof of its validity. Also see
 \cite{clr} for an alternative proof based directly on the rejection-sampling argument, as well as for a numerical illustration.

\section{Coupling Methods}
\label{coupmet}
All algorithms described so far require the coupling of a finite or infinite number of paths in finite time. This is the greater difficulty of applying perfect sampling algorithms to continuous state spaces, especially those with unbounded spaces (which is the case for most routine applications in Bayesian computation) and this is where the greatest ingenuity is required to run perfect sampling in more realistic settings. A good coupling method must be usable in practice and it is even better if it is implementable for different models with the same degree of success. In this section, we review some of the most useful coupling techniques, which essentially belong to two different types: (I) induce a ``common regeneration state" that all sample paths must enter with a positive probability; and (II) explore hidden discretization and hence effectively convert the problem into one with a finite state space.

\subsection{Splitting Technique}\label{sec:split}

A very common technique for coupling MCMC paths is initiated in
\cite{splitting} and discussed in detail by \cite{tho}. Consider the Markov chain $\bo{X}_t$ defined using the transition kernel $K$ and suppose there is a set $C$ (called {\it small set}) and  there
exist $t>0$, $0< \epsilon < 1$, and a \emph{probability measure }$\nu$ such that
$$K_t(x,dy)\ge \epsilon \nu(dy), \; \; \forall\  x\in C,$$
where $K_t$ represents the $t$-step transition kernel. Thus, for any $x\in C$
\beq \label{splitchain}
K_t(x,dy)=\epsilon \nu(dy)+ (1-\epsilon) \frac{K_t(x,dy)-\epsilon \nu(dy) }{ 1-\epsilon}=\epsilon \nu(dy)+(1-\epsilon)Q(x,dy),
\eeq
where $Q(x,dy)=[K_t(x,dy)-\epsilon \nu(dy)] /(1-\epsilon)$. The representation given by (\ref{splitchain}) is important because with probability $\epsilon$ the updating of the chain will be done using the probability measure $\nu$, that is, {\it independently} of the chain's current state. If at time $t$ all the paths are in set $C$ and all the updates use the same random numbers $\xi$ that lead to the transition into the $\nu$ component of (\ref{splitchain}), then all paths will coalesce at time $t+1$, even if there are uncountably many. However, for a set $C \subset \cas$ it will be difficult, if not impossible, to determine whether it contains all paths at a given time. This problem is alleviated in the case of CFTP where the existence of successful coupling has been shown (see \cite{foss}) to be equivalent to the uniform ergodicity of the chain $\bo{X}_t$, in which case the small set is the whole sample space, $\cas$, so all paths are automatically within a small set at all times. An example where this idea has been brought to fruition is the {\it multigamma coupler} introduced by \cite{murdoch-green:continuous}, following the gamma coupler of \cite{lindv}. The method is further developed by \cite{murdoch00exact} in the context of perfect sampling from continuous state distributions.

The multigamma coupler applies when the update kernel density $f(\cdot|x)$ of the Markov chain is known. In addition, it requires that there  is a nonnegative function $r$ such that
\beq
\label{bmg}
f(y|x)\ge r(y), \; \; \forall x,y \in \cas.
\eeq If we denote $\rho=\int r(y)dy>0$, then in line with the splitting technique discussed above we can write
\beq\label{multigamma}
P(X_{t+1}\le y|X_t=x)=\rho R(y) + (1-\rho)Q(y|x),
\eeq
where $R(y)=\rho^{-1} \int_{-\infty}^y r(v)dv$ and $Q(y|x)=(1-\rho)^{-1} \int_{-\infty }^ y [f(v|x)-r(v)]dv$.

As a simple example, assume that the transition kernel has the Gamma density $f(y|a,b_x)= y^{a-1}b_x^a \exp(-yb_x)/\Gamma(a)$, where $a$ is fixed, $b_x$ depends on the previous state $X_t=x$ but it is always
 within a fixed interval, say  $b_x \in [b_0,b_1]$, where $b_0$ and $b_1$ are known constants.
Then we can set  $r(y) =y^{a-1}b_0^a \exp(-yb_1)/\Gamma(a)$,  which yields
$\rho=(b_0/b_1)^a$. At each $t$, we sample $\xi\sim {\rm Bernoulli}(\rho)$, and if $\xi=1$, we draw
$y$ from Gamma($a, b_1$), and let all paths $X_{t+1}=y$ regardless of their previous states, hence coalescence takes place. If $\xi=0$, then we draw from the $Q$ component in (\ref{multigamma}) (though this step requires drawing from a non-standard distribution).

In situations when no uniform bound can be found on $\cas$ for (\ref{bmg}) to hold, \cite{murdoch-green:continuous} propose partitioning $\cas = \cas_1 \cup \ldots \cup  \cas_m$ and bounding the kernel density $f$ on each $\cas_i$ with $r_i$ and introduce a {\it partitioned multigamma coupler} for this setting.  A more difficult coupling strategy has been described in \cite{MR2108860} in the case of geometrically (but not necessarily uniformly) ergodic chains, though the approach has not been implemented on a wider scale.

There is a direct connection  between the multigamma coupler and the Ro-CFTP in Section~\ref{sec:readonce}. With a block of size $K=1$ the multigamma coupler construction implies that the probability of coalescence within the block is $\rho$. As described above, we can therefore sample a geometric $T$ with success probability $\rho$, and start from a coalesced value, namely, an independent draw from $R(y)$ in (\ref{multigamma}). We then run the chain forward for $T-1$ steps conditioning on non-coalesced blocks, namely, we use the $Q$ component of (\ref{multigamma}) as the transition kernel. The resulting value then is an exact draw from $\Pi$ (\cite{murdoch-green:continuous}).

There is also a close connection between the multigamma coupler and the slice sampler (see Section \ref{psls}), as both can be viewed as building upon the following simple idea: for a given  (not necessarily normalized) density $g(y)$, if $(U, Y)$ is distributed uniformly on $\Omega_g=\{(u, y): u\le g(y)\}$, then the marginal density of $Y$ is proportional to $g(y)$. Therefore, when $f(y|x)\ge r(y)$ for all $x$ and $y$, we have
\beq\label{eq:reject}
\Omega_r=\{(u, y): u\le r(y)\} \subset \Omega_{f,x}=\{(u, y): u\le f(y|x)\}, \ \forall\ x\in \cas.
\eeq
For simplicity of illustration, let us assume all $\Omega_{f,x}$ are contained in the unit square $[0,1]\times[0,1]$. Imagine now we use rejection sampling to achieve the uniform sampling on $\Omega_{f,x}$ for a particular $x$ by drawing uniformly on the unit square. The chance that the draw $(u, y)$ will fall into $\Omega_r$ is precisely $\rho$, and more importantly, if it is in $\Omega_r$, it is an acceptable proposal for $f(y|x)$ regardless of the value of $x$ because of (\ref{eq:reject}). This is the geometric interpretation of how the coalescence takes place for splitting coupling, which also hints at the more general idea of coupling via a common proposal, as detailed in the next section.

\subsection{Coupling via a Common Proposal}

The idea of using a common proposal to induce coalescence was given in \cite{breyer-tech} as a way to address the second challenge discussed in Section 1.2.3. (Note however that this strategy does not directly address the first challenge, namely discretizing a continuous set of paths
into a finite set; that challenge is addressed by, for example, the augmentation method described in the next subsection, or by other clever methods such as the multishift coupler in \cite{wilson:multishift}.) Imagine that we have managed to reduce the number of paths to a finite one. In practice, it may still take a long time (possibly {\it too} long) before all paths coalesce into one. Intuitively, one would like to make it easier for paths that are close to each other to coalesce faster.

Remarkably, the description of coupling via a common proposal can be formulated in a general setting irrespective of the transition kernel used for the chain, as long as it has a density.
Suppose the chain of interest has transition kernel with the (conditional) density $f( \cdot |X_{t})$. Instead of always accepting the next state as $X_{t+1} \sim f(\cdot |X_{t})$, we occasionally replace it with a random draw  $\tilde{Y}$  sampled from a user-defined auxiliary density $g$. Thus, the  $X_{t+1}$ from the original chain plays the role of a proposal and is no longer guaranteed to be the next state; we therefore relabel it as $\tilde{X}_{t+1}$.

Instead, given $X_t=x$, the next state $X_{t+1}$ is given by the following updating rule,
\beq
{X}_{t+1}=\left \{
\begin{array}{cc}
\tilde{Y}, &  \; \mbox{ if  }  \; \frac{ f(\tilde{Y}|x)g(\tilde{X}_{t+1})}{ f(\tilde{X}_{t+1}|x) g(\tilde{Y})} > U;\\
\\
\tilde{X}_{t+1}, & \; \mbox{otherwise.}\\
\end{array}
\right .
\label{couprop}
\eeq
where $U \sim \ru(0,1)$ and is independent of any other variables. In other words,
the above coupler makes a choice between two independent random variables
$\tilde{X}_{t+1}$ and $\tilde{Y}$ using a Metropolis-Hastings (M-H) acceptance ratio.  Note that the M-H accept-reject move is introduced here simply to ensure that the next state of the chain has distribution density $f(\cdot |X_t)$ even if occasionally the state is ``proposed" from $g$. The coupling via a common proposal tends to increase the propensity of coalescing
paths that are close to each other.  More precisely, suppose that two of the paths are close, i.e.  $X_t ^{(1)}\approx X_t^{(2)}$. Then the ratios (\ref{couprop}) will tend to be similar for the two chains, which implies that both chains will likely accept/reject $\tilde{Y}$ simultaneously.

It is also worth emphasizing that the above scheme requires a modification in order to be applicable to  Markov chains whose transition kernels' densities are not available (either they do not exist or not easily computable).  For example, for an M-H algorithm, coupling can be done at the level of proposals using an independent auxiliary variable generated from an auxiliary density $g$  as in \cite{breyer-tech}.
 



This perhaps is best seen via a toy example. Suppose our target distribution is $N(0,1)$, and we adopt a random walk Metropolis, that is, the proposal distribution is $q(y|x)=N(y-x)$, where $N(z)$ is the density of $N(0,1)$. Clearly, because $N(z)$ is continuous, two paths started in different points of the sample space will have zero probability of coalescing if we just let them ``walk randomly".  To stimulate coalescence, we follow the ideas in \cite{breyer-tech} and create an intermediate step in which the proposals used in the two processes can be coupled.

\begin{figure}
\begin{center}
\includegraphics[height=4in,angle=0]{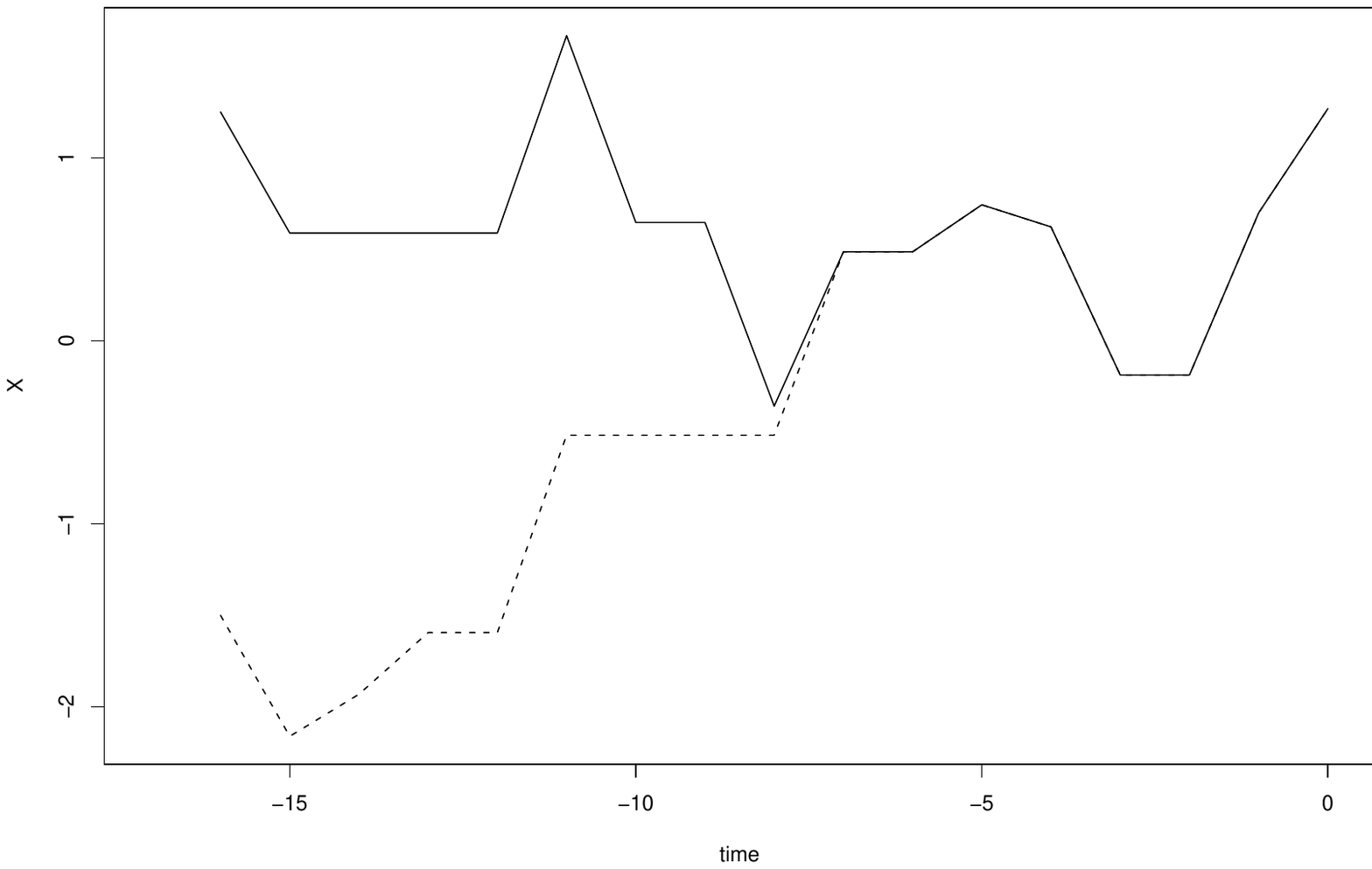}
\caption{{\em Illustration of coupling with proposals for two paths. }}
\label{fig:couprop}
\end{center}
\end{figure}

More precisely, at each time $t$ we sample $\tilde{Z}_{t+1}\sim g(\cdot)$, where $g$ is an auxiliary density supported on $\RR$.  Suppose that the proposal for chain $i$ at time $t$ is $\tilde{Y}_{t+1}^{(i)}$ where $\tilde{Y}_{t+1}^{(i)}\sim N(X_t^{(i)}, 1)$. We then define
\beq
\tilde{W}_{t+1}^{(i)}=\left \{
\begin{array}{cc}
\tilde{Z}_{t+1} , & \mbox{ if }  \frac{N(\tilde{Z}_{t+1}-X_{t}^{(i)})g(\tilde{Y}_{t+1}^{(i)})}{
N(\tilde{Y}_{t+1}^{(i)}-X_t^{(i)})g(\tilde{Z}_{t+1})} > U\\
\\
\tilde{Y}_{t+1}^{(i)}, & \mbox{ otherwise }\\
\end{array} \right. ,
\label{eq:cmh}
\eeq
where $U\sim \ru(0,1)$ is independent of all the other variables. The proposal $\tilde{W}_{t+1}^{(i)}$ is accepted using the usual M-H strategy because its density is still the density of the original proposal, i.e. $N(X_t^{(i)}, 1)$; the next state is then either $\tilde{W}_{t+1}^{(i)}$ (acceptance) or $X_t^{(i)}$ (rejection). What has changed is that regardless of which paths the chains have taken, their M-H proposals now have a positive probability to take on a common value $\tilde{Z}_{t+1}$ for all those chains for which the first inequality in (\ref{eq:cmh}) is satisfied. This does not guarantee coupling but it certainly makes it more likely.  In Figure \ref{fig:couprop} we show two paths simulated using the simple model described above, where the two paths first came very close at $t=-8$ and then they coalesced at $t=-7$.

\subsection{Coupling via Discrete Data Augmentation}
\label{coup-aug}
Data augmentation (\cite{Tanner:1996fk}), known also as auxiliary variable method in statistical physics,
is a very effective method for constructing efficient MCMC algorithms; see \cite{van} for a review. It turns out to be useful for perfect sampling as well, because we can purposely consider  auxiliary variables that are discrete and therefore convenient for assessing coalescence. Specifically, suppose our target density is $f(x)$, where $x$ may be continuous. Suppose we have a way to augment $f(x)$ into $f(x,l)$, where $l$ is discrete.
If we can perform Gibbs sampling via $f(x|l)$ and $f(l|x)$, then we automatically will have a Markov sub-chain with $f(l)$ as the stationary density (note the sub-chain with $l$ only is Markovian because the Gibbs sampler here only involves two steps). Therefore, we have effectively turned the continuous problem for $f(x)$ into a discrete one because once we have an authentic draw from $f(l)$, then we can easily get a corresponding authentic draw from $f(x)$ by sampling from $f(x|l)$.

To illustrate, consider finite mixtures, where the obvious auxiliary variable
is the indicator variable indicating the mixture component from which a sample is obtained.
The coupling via augmentation has been successfully implemented  by
\cite{hobert-robert-titterington:mixtures} in the case of two-component mixtures of distributions and by \cite{murdoch-towards} in the case of Bayesian mixture priors.
Below is one of the examples discussed by \cite{hobert-robert-titterington:mixtures}, which we recast in order  to crystalize the essence of discrete data augmentation.

 Consider the mixture  $\alpha f_0(d)+(1-\alpha)f_1(d)$, where only the mixture proportion
 $\alpha$ is unknown
 and therefore we seek its posterior density, assuming a uniform prior on $(0,1)$.
 Given a sample $\{ d_1,\ldots, d_n \}$ from the mixture, the posterior for $\alpha$ is proportional to
 \begin{equation}\label{eq:mixt}
 p(\alpha|\vec d)\propto \prod \limits _{i=1}^n \{ \alpha f_0(d_i)+(1-\alpha)f_1(d_i)\},
 \end{equation}
 involving $2^n$ terms
 when expanded; note that here we use $\vec d=\{d_1,\ldots, d_n\}$ to denote the data instead of the
 original $\{x_1,\ldots, x_n\}$, to avoid the potential
 confusion of our generic notation that uses $X$ for the sampling variable, which is $\alpha$ here.
 Let the latent variables $\vec{z}=\{ z_1,...,z_n\}$ be such that $z_i=0$ if
 $d_i$ has been generated from $f_0$ and $z_i=1$ if $d_i$ has been generated from $f_1$.
 Then it is easy to see that
 \begin{equation}\label{eq:drawz}
  P(z_i=1|\vec{d},\alpha)=\frac{(1-\alpha) f_1(d_i)}{\alpha f_0(d_i)+(1-\alpha)f_1(d_i)}:=p_i
  \end{equation}
  and
  \begin{equation}\label{eq:drawp}
  P(\alpha|\vec{z})={\rm Beta} \left(n+1-\sum_{i=1}^n z_i, \sum_{i=1}^n z_i +1 \right).
  \end{equation}
This implies that we can construct the discrete augmentation as $l=\sum_i z_i$, which has
a non-homogenous binomial (NhB) distribution NhB$(n, \vec p)$, where $\vec p=\{p_1, \ldots, p_n\}$.
That is, $l$ is the sum of $n$ independent but not necessarily identically distributed Bernoulli variables. Given this data augmentation scheme $f(\alpha, l)$, the algorithm given in \cite{hobert-robert-titterington:mixtures} can be reformulated as follows.
  \begin{enumerate}
  \item Because of (\ref{eq:drawp}), given $l_{t}=l$, we generate $\alpha_{t+1} \sim {\rm Beta}(n+1-l, l+1)$,
  which can be accomplished by drawing $w_j \sim
  {\rm Exponential}(1) \mbox{ for } j\in \{1,\ldots,n+2\}$ and then letting
  \begin{equation}\label{eq:psi}
  \alpha_{t+1}
  =\frac{\sum \limits _{i=1}^{n+1-l} w_i}{\sum \limits _{i=1}^{n+2} w_i}.
  \end{equation}
  \item Given $\alpha_{t+1}=\alpha$, because of (\ref{eq:drawz}), we need to draw $l_{t+1}$ from NhB$(n, \vec{p}(\alpha))$, where $\vec{p}(\alpha)=\{p_1, \ldots, p_n\}$, with
      $p_i\equiv p_i(\alpha)$ given by the right-hand side of (\ref{eq:drawz}).
      This draw is accomplished by generating independent $u_i\sim {\rm Uniform}(0,1)$ and letting \begin{equation}\label{eq:drawl}
      l_{t+1}=\sum_{i=1}^n \one\{u_i \le p_i\},
      \end{equation}
      where $\one\{A\}$ is the usual indicator function of event $A$.
\end{enumerate}

Combining (\ref{eq:psi})-(\ref{eq:drawl}), we see that the SRS from $l_t$ to $l_{t+1}$ can be written as
\begin{equation}\label{eq:srs}
l_{t+1}\equiv \phi(l_t; \vec{u}, \vec{w})=\sum_{i=1}^n \one\left\{u_i \le \left[1+ \left(\frac{\sum_{i=1}^{n+2} w_i}{\sum_{i=1}^{n+1-l_t}w_i} - 1\right)^{-1}\frac{f_0(d_i)}{f_1(d_i)}\right]^{-1}\right\}.
\end{equation}
For given $\vec{u}=\{u_1,\ldots, u_n\}$ and $\vec{w}=\{w_1,\ldots, w_n\}$, the function $\phi$ in (\ref{eq:srs}) is evidently  increasing in $l_t$ and thus defines, with respect to the natural integer ordering,  a monotone Markov chain on the state space $S_l=\{0,\ldots, n\}$, with the ceiling and floor states given by $l=0$ and $l=n$. Through data augmentation we therefore have converted the problem of drawing from the continuous distribution given by (\ref{eq:mixt}) to one in which the sample space is the finite discrete space $S_l$, given by (\ref{eq:srs}), for which we only need to trace the two extreme paths starting from $l=0$ and $l=n$.

\subsection{Perfect Slice Sampling}
\label{psls}

Slice sampling is based on the simple observation that sampling from $\Pi$ (assumed to have density $\pi$) is equivalent to sampling from the uniform distribution
$g(u,x) \propto 1_{\{u\le f(x)\}}$ where $f$ is {an} un-normalized version of $\pi$ and is assumed known.  One can easily see that the marginal distribution
of $x$ is then the desired one. In turn, the sampling from $g$ can be performed using a Gibbs  scan in which both steps involve sampling from uniform distributions:
 \begin{description}
 \item[Step I] Given $X_t$, sample $U \sim \ru(0,f(X_t))$;
   \item[Step II] Given $U$ from Step I, sample $X_{t+1}\sim \ru[A(U)]$, where $A(w)=\{y: f(y)\ge w\}$.
   \end{description}
   Here, for simplicity, we assume $A(U)$ has finite Lebesgue measure for any $U$; more general implementations of the slice sampler are discussed in \cite{wake} and  \cite{MR1994729}.  The coupling for slice sampling has been designed by
  \cite{MR1858405} under the assumption that there exists a minimal element $x_{\min} \in \cas$  with respect with the order $x\prec y \Leftrightarrow f(x) \le f(y)$.

  Specifically, the {\it perfect slice sampler} achieves coupling via introducing common random numbers into the implementation of Steps I-II in the following fashion. For Step I, regardless of the value of $X_t$, we implement by drawing $\epsilon \sim \ru(0,1)$ and then letting $U=U(X_t)=\epsilon f(X_t)$; hence all $U(X_t)$'s share the same random number $\epsilon$.

  Given the $U=U(X_t)$ from Step I, we need to implement Step II in such a way that there is a positive (and hopefully large) probability that all $X_{t+1}$ will take the same value regardless of the value $X_t$. This is achieved by forming a sequence of random variables $\bW=\{W_j \}_{ j=1,2,\ldots}$, where $W_1 \sim \ru[A(f(x_{\min}))]$ and $W_{j} \sim \ru[A(f(W_ {j-1}))]$, for any $j \ge 2$. The desired draw $X_{t+1}$ is then the first $W_j \in A(U(X_t))=A(\epsilon f(X_t))$, that is,
  \beq\label{eq:ss}
  X_{t+1}\equiv \phi(X_t,(\epsilon, \bW))=W_{\tau(X_t)},
  \eeq
  where $\tau(x)=\inf\{j: f(W_j) \ge \epsilon f(x)\}$.

  In \cite{MR1858405} it is proven  that, almost surely, only a finite number of the elements of the sequence $\bW$ are needed in order to determine $\tau(x)$.  The correctness of the algorithm is satisfied if $W_{\tau(x)} \sim \ru [A(\epsilon f(x))]$, and in \cite{MR1858405} this is established by viewing it as a special case of adaptive rejection sampling. Here we provide a simple direct proof. For any given $x$, denote $A^{(x)}=A(\epsilon f(x))$, and $B_j^{(x)}=\{(W_1,\ldots, W_j): f(W_i)<\epsilon f(x),
  i=1,\ldots, j\}$. Then clearly for any $k\ge 1$, $\{\tau(x)=k\}=\{W_k\in A^{(x)}\}\cap B_{k-1}^{(x)}$ (assume $B_0^{(x)}=\cas$ for any $x \in \cas$). Hence, for any (measurable) set $C\subset A^{(x)}$, we have
\begin{eqnarray}
P(\{W_{\tau(x)}\in C\}|\tau(x)=k)&=& \frac{P(\{W_k\in C\cap A^{(x)}\}\cap B_{k-1}^{(x)})}{P(\{W_k\in A^{(x)}\}\cap B_{k-1}^{(x)})} \nonumber \\ \nonumber
&=&
\frac{E\left[E\left(\one\{W_k\in C\}\one\{B_{k-1}^{(x)}\}|W_1,\ldots, W_{k-1}\right)\right]}{E\left[E\left(\one\{W_k\in A^{(x)}\} \one\{B_{k-1}^{(x)}\}|W_1,\ldots, W_{k-1}\right)\right]} \\
&=&
\frac{E\left[\one\{B_{k-1}^{(x)}\}P(\{W_k\in C\}|W_{k-1})\right]}{E\left[\one\{B_{k-1}^{(x)}\}P(\{W_k\in A^{(x)}\}|W_{k-1})\right]}. \label{eq:resu}
\end{eqnarray}
In the above derivation, we have used the fact that $\{W_1,\ldots, W_k\}$ forms a Markov chain itself. Given $W_{k-1}=w$, $W_k$ is uniform on $A(f(w))$ by construction, so $P(\{W_k\in B\}|W_{k-1}=w)=\mu(B)/\mu(A(f(w)))$, where $\mu$ is the Lebesgue measure. Consequently, the last ratio in (\ref{eq:resu}) is exactly $\mu(C)/\mu(A^{(x)})$, the uniform measure
on $A^{(x)}$. It follows immediately that $W_{\tau(x)}\sim \ru(A^{(x)})=\ru[A(\epsilon f(x))]$.

\begin{figure}
\begin{center}
\includegraphics[width=3.5in]{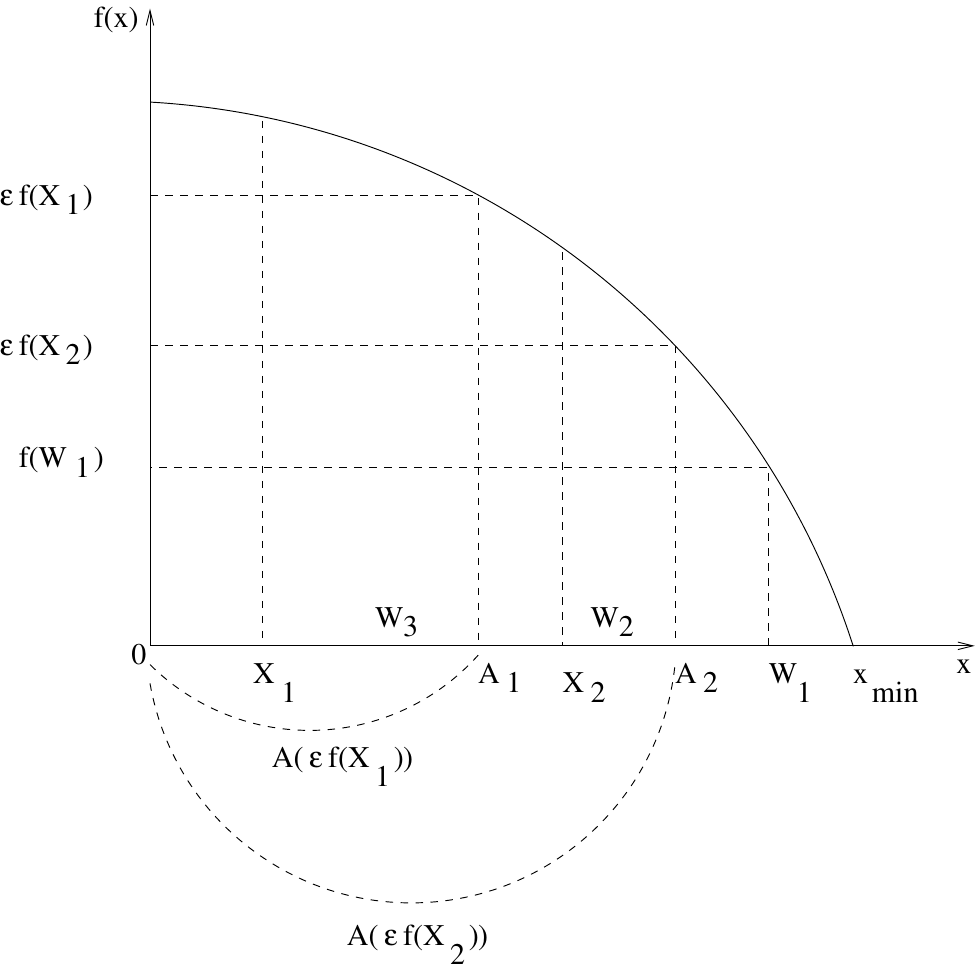}
\caption{{\em Illustration of perfect slice sampling.}}
\label{fig:slice}
\end{center}
\end{figure}

 To visualize how Steps I-II achieve coupling, Figure \ref{fig:slice} depicts the update for two paths in the simple case in which $f$ is strictly decreasing with support $(0,x_{\min})$. Suppose the two chains are currently in $X_1$ and $X_2$. Given the $\epsilon$ drawn in  Step I, the monotonicity of $f$ allows us to write $A(\epsilon f(X_1))=(0,A_1)$ and $A(\epsilon f(X_2))=(0,A_2)$. Step II then starts by sampling $W_1 \sim \ru(0,x_{\min})$ and, since it is not in either of the intervals $(0,A_1)$ or $(0,A_2)$, we follow by sampling uniformly
$W_2\sim \ru(0,W_1)$ which is the same as sampling $W_2 \sim \ru [A(f(W_1))]$ since $f$ is decreasing.
Because  $W_2 \in (0, A_2)$, we have $\tau(X_2)=2$ so $X_2$ is updated into $W_2$.
As $W_2 \not \in (0,A_1)$ we continue by sampling $W_3 \sim \ru(0,W_2)$ and since $W_3 \in (0,A_1)$ we can set $\tau(X_1)=3$. Thus, in the case illustrated by Figure \ref{fig:slice} the updates are
$\phi(X_1,\bW)=W_3$ and $\phi(X_2,\bW)=W_2$. To understand why this construction creates opportunity for coupling, imagine that the second uniform draw, $W_2$, happens to be smaller than $A_1$. In this case, $\tau(X_1)=\tau(X_2)=2$ so both $X_1$ and $X_2$ are updated into $W_2$ which means that the two paths have coalesced. In fact, for all $X \in (0,x_{\min})$ with the property that $ f(X) \le f(W_1)/\epsilon$ we have $\phi(X, \bW)=W_1$ and for all $X$ such that $ f(W_1)/ \epsilon < f(X) \le  f(W_2)/\epsilon$, $\phi(X, \bW)=W_2$, and so on. This shows how the continuous set of paths  is discretized in only one update.

Figure \ref{fig:slice} also illustrates that the density ordering $X_2\prec X_1$ (since $f(X_2)< f(X_1)$) is consistent with the same ordering for the updates: $\phi(X_2,\bW)=W_2 \prec \phi(X_1,\bW)=W_3$ because $f(W_2)\le f(W_3)$ by construction. This is true in general because if $X_2\prec X_1$, that is $f(X_2)\le f(X_1)$, then $\tau(X_2)\le \tau(X_1)$ because $A(\epsilon f(X_1))\subset A(\epsilon f(X_2))$.
Consequently, $W_{\tau(X_2)}\prec W_{\tau(X_1)}$.
This property implies that we can implement the monotone CFTP as described in Section 1.3.1,
 because a maximal $x_{\max}=0$ exists in this case.  In situations in which the extremal states cannot be found,  \cite{MR1858405} show how to construct bounding processes for this perfect slice sampler.

 \section{Swindles}
\label{swind}
 The term swindle has traditionally been used in Monte Carlo literature to characterize any
  strategy or modification that either reduces the computational effort or increases the efficiency of the algorithm (e.g.,
\cite{Simo:76:CSS} and \cite{Gentle:98:SS}). Usually, swindles are relatively easy-to-implement generic methods applicable to a wide class of algorithms.
In the following we describe some of the swindles proposed that are either for or taking advantage of perfect sampling ideas.

\subsection{Integrating Exact and Approximate MCMC Algorithms}

It is probably clear by now to the statistician with some travel experience in the MCMC kingdom that perfect sampling may not be the vehicle that one could take on every trip. But it is possible to extend its range considerably if we couple it with more traditional MCMC methods. Here we describe such an approach devised by \cite{MR2278096}
to deal with Bayesian computation in cases where the sampling density is known only up to a constant that depends on the model parameter, and hence the likelihood function itself cannot be evaluated directly.

More precisely, consider the case in which the target of interest is the posterior density
$\pi(\theta|y)\propto p(\theta)p(y|\theta)$, where $p(\theta)$ is the prior density and $p(y|\theta)$ is the sampling density of the data. There is a large spectrum of problems, e.g.,  Markov random fields, image analysis, Markov point processes, Gaussian graphical models, neural networks, for which $p(y|\theta)$ is known only up to a constant, i.e.,\
$p(y|\theta)=q(y|\theta)/C_{\theta}$, with the functional form of $q$ known but the normalizing constant $C_{\theta}$ unknown, in the sense that its value at any particular $\theta$ is hard or even impossible to calculate. Obviously, for such problems, the classical MCMC approach cannot be directly implemented. For instance, a Metropolis algorithm with a symmetric proposal, moving from $\theta \rightarrow \theta'$, would require the calculation of the acceptance ratio
\beq
\label{mhratio}
\alpha(\theta';\theta)=\min\left \{ 1, \frac{p(\theta') q(y|\theta')}{p(\theta) q(y|\theta) } \times \frac{C_{\theta} }{ C_{\theta'}} \right\}
\eeq
 which involves the unknown ratio of two normalizing constants, $C_\theta/C_{\theta'}$, a problem which occurs in many areas (see for example \cite{mengwong96}, \cite{MR1647507}, and \cite{mengshill}).

 One obvious way to deal with this problem is to use Monte Carlo or other approximations to estimate each ratio needed in the implementation of Metropolis-Hastings algorithm. A more creative and ``exact" solution is proposed by \cite{MR2278096} with the help of perfect sampling. The idea is to add into the mix an auxiliary variable $x$ such that the chain updates not only $\theta$ but $(\theta,x)$ via MH sampling with an acceptance ratio in which no unknown constant appears. Since the auxiliary variable is just a computational artifact, as long as the marginal distribution of $\theta$ is preserved there is a lot of freedom in choosing how to update $x$. In particular, we consider updating $(\theta, x)$ via a proposal $(\theta', x')$ in which the proposal 
 $\theta'$ is generated as in the original chain (does not depend on $x$) but
 $x'|\theta',\theta, x \sim q(\cdot|\theta')/C_{\theta'}$. Essentially, $x'$ is pseudo-data simulated from the sampling distribution when the parameter is equal to the proposal, $\theta'$.  For the new chain, the acceptance ratio is then
 \begin{equation}\label{eq:moller}
 \tilde{\alpha}=\min\left \{ 1, \frac{p(\theta') q(y|\theta') q(x|\theta)}{ p(\theta) q(y|\theta) q(x'|\theta')} \right\},
 \end{equation}
 which no longer involves any unknown normalizing constant.

 A perceptive reader may have immediately realized that the above scheme simply transfers one difficult problem into another, namely, simulating from the original sampling density $p(\cdot|\theta')=q(\cdot|\theta')/C_{\theta'}$. Since $C_{\theta'}$ is not available, direct methods such as inverse CDF are out of question (even when they are applicable otherwise). We can of course apply Metropolis-Hastings algorithm itself for this sampling, which will not require any value of $C_{\theta}$ (since here we sample for $x$, not $\theta$). But then we would need to introduce a new proposal, and more critically we would need to worry about the convergence of this imbedded Metropolis-Hastings algorithm \emph{within each step} of creating a proposal $(\theta', x')$ as called for by (\ref{eq:moller}). This is clearly cumbersome and indeed entirely defeats the purpose of introducing $x'$ in order to have a ``clean" solution to the problem without invoking any approximation (beyond the original Metropolis-Hastings algorithm for $\theta$). This is where the perfect sampling methodologies kick in, because if we have an exact draw from $p(x'|\theta')$, then the acceptance ratio given in (\ref{eq:moller}) is exactly correct for implementing the Metropolis-Hastings algorithm for drawing $(\theta, x)$ and hence for $\theta$. This is particularly fitting, since intractable likelihoods are common in inference for point processes and this is also the area where exact sampling has been most successful.  For instance,  in \cite{MR2278096}, the method is illustrated on the well-known Ising model  which  was proposed as a main application in Propp and Wilson's landmark paper \cite{propp-wilson:exact-sampling}, which is a ``must" for any tourist of the magic land of perfect sampling. 
 
{We emphasize} that the method discussed here is only one among a number of promising attempts that have been made to couple the power of traditional MCMC with the precision of perfect sampling such as in  \cite{eff-cftp} and \cite{murr-gha}. See also \cite{murr-thesis} for related ideas and algorithms. 

 \section{Unbiased MCMC: A fertile trade-off?}

Perfect sampling's promise of drawing iid samples from the target has, for the practical statistician, important consequences as it eliminates the  bias in estimation and, perhaps most importantly, it dissolves the need for computing (or bounding) the distance between the distribution  of the chain at $t$-th update and the target. However, the original promise has remained mostly aspirational because perfect sampling remains to this day limited in its application to statistical computation. Nevertheless, recent developments on unbiased MC--- see Chapter ??? in this handbook for by Yves Atchade and Pierre Jacob for a full treatment---have demonstrated that coupling can be used  more widely and, arguably, much more effectively as long as one is willing to promise less. This line of work builds on pioneering ideas of \cite{glynn2014exact} and \cite{glynn2016exact}, who put forth strategies for exact estimation of integrals via Markov chain Monte Carlo, forming the class of so-called unbiased MCMC (UMCMC). 

Specifically, the L-lag coupling of \cite{biswas2019estimating} and \cite{jacob2020unbiased}  delivers: i) unbiased estimators of $\E[h(X_\pi)]$ for any (integrable) $h$, where $X_\pi$ denote a random variable defined by $\pi(X)$ and ii) a bound on the total variation distance between $\pi_k$, the distribution of $X_k$, and $\pi$.  This immediately recovers   the two  practical consequences of perfect sampling we mentioned above. Switching from being able to perform iid sampling from the target to unbiased estimation relies on an apparent trade-off: instead of coupling  chains started from all possible values in the state space, the UMCMC strategy is to couple only two copies of the MCMC chain of interest. The two chains, say ${\cal X}=\{X_t, t\ge 0\}$ and ${\cal Y} = \{Y_t, t\ge 0\}$  are started from the same initial distribution, are updated using the same transition kernel, and  are shifted by a time lag $L$. 

Specifically, the two chains ${\cal X}, {\cal Y}$ are coupled so that there exists with probability one a finite stopping time $\tau_L$ such that $X_{t}=Y_{t-L}$ for all $t \ge \tau_L$.  This construction allows \cite{jacob2020unbiased} (for $L=1$) and \cite{biswas2019estimating} (general $L$) to show that the following estimator based on both $\cal X$ and $\cal Y$,
\beqn
H_{k,L}({\cal X,Y}) = h(X_k)+\sum_{j=1}^{J_{k,L}} \left[h(X_{k+jL})-h(Y_{k+(j-1)L})\right], 
\label{eq:hkLf}
\eeqn
where $J_{k,L}=\max\left\{0,  \lceil \frac{\tau_L-L-k}{L} \rceil \right\}$, 
is an unbiased estimator for $\E[h(X_\pi)]$ for any $k\ge 0$ (under mild conditions). This follows from noticing that (\ref{eq:hkLf}) is the same as $\sum_{j=1}^\infty[h(X_{k+jL})-h(Y_{k+(j-1)L})]$ because the terms  corresponding to $j \ge J_{k,L}+1$ are zero due to the coupling scheme.  Furthermore, for the purpose of calculating expectations, we can replace $h(Y_{k+jL})$
by $h(X_{k+jL})$ for any $j\ge 0$ because they have identical distributions by construction.  However, $h(X_k)+\sum_{j=1}^\infty[h(X_{k+jL})-h(X_{k+(j-1)L})]$ is nothing but $\lim_{t\rightarrow \infty} h(X_t)$, which has the same distribution as $h(X_\pi)$. 

In our discussion (\cite{cmdisc}) of 
\cite{jacob2020unbiased} we have identified a simple way to reduce the variance of $H_k$ via control variates. Specifically, after rearranging the terms in \eqref{eq:hkLf}
we find an alternative form for $H_k$,
\beqn\label{eq:hkLb}
H_{k,L}({\cal X,Y}) 
= h(X_{k+LJ_{k,L}}) + \sum_{j=0}^{J_{k,L}-1} \left[h(X_{k+jL})-h(Y_{k+jL})\right].
\eeqn
that spotlights the time-backwards correction term 
\beq
\sum_{j=0}^{J_{k,L}-1} \left[h(X_{k+jL})-h(Y_{k+jL})\right]
\label{cv1}
\eeq 
which has known mean zero (conditional on $\tau$), thus making it a natural candidate for a control variate.

Moreover, \cite{biswas2019estimating} have shown that (under mild regularity conditions) the total variation distance between $\pi_k$, the distribution of $X_k$, and $\pi$ is  bounded by a very simple function of $\tau_{L}$ and $(k, L)$:
\beq\label{eq:biswas}
d_{\rm TV}(\pi_k, \pi) \le \E[J_{k,L}].  
\eeq
In \cite{meng2022} we explore the use of \eqref{cv1} to reduce the  variance of $H_{k,L}$ in \eqref{eq:hkLf}. Setting, $\Delta_{k,j} = \left[h(X_{k+jL})-h(Y_{k+jL})\right ]$, in \cite{meng2022} we consider control variates of the form $\sum_j \eta_j \Delta_{k,j}$ and derive the optimal choice for the coefficients $\eta_j$. The optimality criteria is set in terms of sharpening the TV distance between $\pi_k$ and $\pi$ and leads to 
$\eta_j = \one_{\{S_j>0.5\}}$ where $S_j=\Pr(\tilde J_{k,L}\ge j)=\Pr(J_{k,L}>j)+0.5\Pr(J_{k,L}=j), \quad {\rm for\ any}\ j \ge 0$. 
The resulting bound  is always tighter than the total variation bound \eqref{eq:biswas} but difference is of practical significance only when $\tau_L$ has a small median - we refer the reader to  \cite{meng2022} for additional details and examples.

\section{Where Are The Applications?}

 The most resounding successes of  perfect sampling have been reported from applications involving
 finite state spaces, especially in statistical physics (e.g. \cite{propp-wilson:exact-sampling}, \cite{sw-hub}
\cite{haggstrom-nelander:antimonotone},  \cite{serv}) and  point processes   (e.g. \cite{MR1788098},
 \cite{MR2156552}, \cite{MR2015032}, \cite{MR1673122}, \cite{MR1699662}, \cite{kend}, \cite{MR2004226}, \cite{MR1704559},
 \cite{MR1978833}).  Other applications include sampling from truncated distributions (e.g., \cite{philrob},  \cite{beskos}),
 queuing (\cite{mur-que}) , Bayesian inference (as in \cite{murdoch-green:continuous},  \cite{murdoch-towards}, \cite{moller-nicholls:perfect-tempering},\cite{mharan})  and mixture  of distributions (see
 \cite{hobert-robert-titterington:mixtures},
 \cite{casmenrobtitt}). 
{In the first edition of this chapter,} we commented on the trade-off between perfection and practicality writing that: \begin{quote}
 ``The price one pays for this mathematical precision is that any perfect sampling method refuses to produce a draw unless it is absolutely perfect, much like a craftsman reputed for his fixation with perfection refuses to sell a product unless it is 100\%  flawless. In contrast, any ``non-perfect" MCMC method can sell plenty of its ``products", but it will either ask the consumers to blindly trust their qualities or leave the consumers to determine their qualities at their own risk. The perfect sampling versus non-perfect sampling is therefore a trade-off between quality and quantity.  Like with anything else in life, perhaps the future lies in finding a sensible balance."
 \end{quote}
The emergence of UMCMC is one such instance of a sensible balance. We believe that more research will be devoted to improve its by-products either through ingenious coupling techniques or via unbiased estimators with superior properties. We also hope that, just as we did not anticipate the progress brought on by UMCMC, the future will continue to surprise us and the ideas put forth by the ambitious project that is perfect sampling will deliver other methods to strike that elusive perfect balance. 
 
 

\section*{Acknowledgments}

RVC has been partially supported by the Natural Sciences and Engineering Research Council of Canada and XLM partially by the National Science Foundation of USA.

\def\IfNoDef#1{\ifx #1\undefined \def #1} \IfNoDef \FileName#1{{\tt #1}}\fi
  \IfNoDef \toenglish#1\endtoenglish{[{\em English\/}: #1\unskip]}\fi \IfNoDef
  \Language#1{{\rm (in #1)}}\fi \IfNoDef \RevLanguage#1{{\rm (in #1)}}\fi
  \IfNoDef \AuthoredBy#1{by #1}\fi \IfNoDef \EditedBy#1{\unskip, edited by
  #1}\fi \IfNoDef \Auth{(Author)}\fi \IfNoDef \Ed{(Editor)}\fi \IfNoDef
  \Rev{(Reviewer)}\fi \IfNoDef \Transl{(Translator)}\fi \IfNoDef
  \Ack{Acknowledgement}\fi \IfNoDef \Add{Addendum}\fi \IfNoDef \ComRep{Comment
  and reply}\fi \IfNoDef \Com{Comment}\fi \IfNoDef \Corr{Correction}\fi
  \IfNoDef \Disc{Discussion}\fi \IfNoDef \Pkg{Package}\fi \IfNoDef
  \Ref{Reference}\fi \IfNoDef \AKwd#1{[#1]}\fi
  \IfNoDef\Keywords#1{\begin{quote} Keywords: #1\end{quote}}\fi
  \ifx\Abstract\undefined \long\def\Abstract #1{\begin{quotation}\noindent
  Abstract: #1\end{quotation}}\fi \ifx\BookReview\undefined
  \long\def\BookReview #1{\begin{quotation}\noindent Review:
  #1\end{quotation}}\fi
  \IfNoDef\BeginBookReviewRefs{\begin{center}References\end{center} } { \bgroup
  \advance \parskip 3pt plus 1pt minus .5pt \parindent = 0pt}\fi
  \IfNoDef\EndBookReviewRefs{\par\egroup}\fi
  \IfNoDef\bibline{\leavevmode\raise.5ex\hbox to26pt{\hrulefill\hskip2pt}}\fi
  \let\UC=\uppercase \long\def\D#1{#1} \def\NoPeriod{\unskip\unskip
  \futurelet\next\NoPeriodA} \def\NoPeriodA{\ifx.\next \let\next=\Gobble
  \else\let\next=\relax \fi\next} \def\Gobble#1{}

\bibliographystyle{abbrv}

\end{document}